\documentclass{article}

\usepackage{microtype}
\usepackage{graphicx}
\usepackage{subfig}
\usepackage{booktabs} 
\usepackage{balance} 
\usepackage{subcaption}
\usepackage[square,numbers]{natbib}
\usepackage[utf8]{inputenc} 
\usepackage[T1]{fontenc}    
\usepackage{url}            
\usepackage{booktabs}       
\usepackage{nicefrac}       
\usepackage{xcolor}         
\usepackage[algo2e,noend]{algorithm2e}
\usepackage{booktabs} 

\usepackage{hyperref}

\usepackage[accepted]{icml2024}

\usepackage{amsmath}
\usepackage{amssymb}
\usepackage{mathtools}
\usepackage{amsthm}
\usepackage{amsfonts}       

\usepackage[capitalize,noabbrev]{cleveref}

\theoremstyle{plain}
\newtheorem{theorem}{Theorem}[section]

\theoremstyle{definition}

\theoremstyle{remark}

\usepackage[textsize=tiny]{todonotes}

\title{Population-aware Online Mirror Descent for Mean-Field Games by Deep Reinforcement Learning}
\date{}

\author{%
    Zida Wu$^{1}$,
    Mathieu Laurière$^{2}$,
    Samuel Jia Cong Chua$^{1}$,
    Matthieu Geist$^{3}$,
    Olivier Pietquin$^{3}$,
    Ankur Mehta$^{1}$%
  }
  \renewcommand\footnotemark{}
\date{
  $^1$UCLA 
  $^2$NYU Shanghai
  \\
  $^3$Cohere
}

\begin{document}

\maketitle

\begin{abstract}
Mean Field Games (MFGs) have the ability to handle large-scale multi-agent systems, but learning Nash equilibria in MFGs remains a challenging task. In this paper, we propose a deep reinforcement learning (DRL) algorithm that achieves population-dependent Nash equilibrium without the need for averaging or sampling from history, inspired by Munchausen RL and Online Mirror Descent. Through the design of an additional inner-loop replay buffer, the agents can effectively learn to achieve Nash equilibrium from any distribution, mitigating catastrophic forgetting. The resulting policy can be applied to various initial distributions. Numerical experiments on four canonical examples demonstrate our algorithm has better convergence properties than SOTA algorithms, in particular a DRL version of Fictitious Play for population-dependent policies.
\end{abstract}

\section{Introduction}
\label{sec:introduction}

Multi-agent systems (MAS) \citep{dorri2018multi} are prevalent in real-life scenarios involving a large number of players, such as flocking \citep{olfati2006flocking,cucker2007emergent}, traffic flow \citep{burmeister1997application}, and swarm robotics \citep{chung2018survey}, among others. The study of MAS has garnered significant attention throughout history. Although many models are pure dynamical systems, some try to incorporate elements of rationality using the framework of game theory. As the number of players increases in these multi-agent systems, scalability becomes a challenge \citep{lowe2017multi,rashid2020monotonic}. However, under symmetry and homogeneity assumptions, mean field approximations offer an effective approach for modeling population behaviors and learning decentralized policies which do not suffer from issues of the curse of dimensionality and non-stationarity.

Mean field games (MFGs) \citep{lasry2007mean,huang2007large,carmona2018probabilistic,bensoussan2013mean} provide a framework for large-population games  where agents are identical in their behaviors (policy) and only interact through the distribution of all agents. This implies that, as the number of agents grows, the influence of an individual agent becomes negligible, reducing the interactions among agents to that between a representative individual and the population distribution. The main solution concept in MFGs corresponds to a Nash equilibrium, which represents the situation where no player has an incentive to unilaterally deviate from its current policy. Recently, several learning methods have been proposed to solve MFGs;  see e.g.~\citep{lauriere2022learning} for a survey. The most basic one relies on fixed point iterations, which amounts to iteratively updating the policy of a player and the mean field. However, convergence of Banach-Picard fixed point iterations relies on a strict contraction condition \citep{li2019efficient, guo2019learning}. This condition necessitates Lipschitz continuity with sufficiently small Lipschitz constants, which often fails to hold~\citep{cui2021approximately, anahtarci2023q}.

To address this limitation, several approaches have been proposed, usually based on some form of smoothing. A popular method is Fictitious Play (FP). FP was originally introduced for finite player games~\citep{brown1951iterative,robinson1951iterative,berger2007brown} and has been extended to the MFG setting~\citep{cardaliaguet2017learning,hadikhanloo2019finite,perrin2020fictitious} (see Algo.~\ref{algorithm2} in Appendix). In essence, FP smoothens the mean field updates by averaging historical distributions from past iterations. Notably, convergence of FP can be proved without relying on strict contraction properties, using instead a structural assumption called Lasry-Lions monotonicity. However, FP faces some computational challenges. In particular, it requires averaging policies, which is not a trivial task when these are represented by non-linear approximators such as deep neural networks. Although this difficulty has been successfully tackled in~\cite{lauriere2022scalable} by introducing a variant of FP, fundamental issues remain for FP-based algorithms. First, these algorithms require computing at each iteration the best response against the distribution of the remaining population, which is costly for complex problems. Second, the mean field is updated by averaging uniformly over all past mean fields, which means that the update rate decays with the iteration number and hence progress is very slow after even a moderately large number of iterations. 

These two issues are avoided by the Online Mirror Descent (OMD) method, which was designed for optimization problems, see e.g.~\citep{shalev2012online}, and has been extended to the MFG framework in~\citep{hadikhanloo2017learning,hadikhanloo2018learning,perolat2021scaling} (see Algo. \ref{algorithm3} in Appendix). Compared with fixed-point based algorithms, OMD aims to stabilize the learning process by using past iterations. But in contrast with FP, it aggregates past Q-functions instead of past policies. Here again, convergence can be proved under monotonicity assumption without using a strict contraction. The two major differences with FP-based methods are that: (1) at each iteration, OMD simply requires \emph{evaluating} a policy (instead of computing an \emph{optimal} one) and (2) the update rate remains \emph{constant}. 
Summing Q-functions is not easy when these are approximated using deep networks, but \cite {lauriere2022scalable} proposed an adaptation to a deep RL setting by implicitly learning the sum of Q-functions with the so-called Munchausen trick~\cite{vieillard2020munchausen}.  

The aforementioned works -- and, in fact, most of the MFG literature -- compute the Nash equilibrium for a single population distribution (or distribution sequence for time-dependent problems). In such cases, it is sufficient to consider policies that are myopic to the population distribution and depend only on the individual state. While this is fine in specific scenarios, this assumption restricts the applicability of the MFG theory. Indeed, in general, one would like to know the value function and the equilibrium policy of the players for \emph{any} mean field, i.e., as functions of the population distribution. The \emph{Master equation} has been introduced by Lions to characterize the mean-field dependent value function, see~\citep{cardaliaguet2019master}. The closely related notion of \emph{master policy}, which is a population-dependent policy, has been introduced in~\citet{Perrin2022General} to enable the attainment of Nash equilibrium from any initial distribution. The authors have proposed a method for learning this policy using a deep RL extension of FP for population-dependent policies. While efficient in small scale problems, this method suffers from the aforementioned intrinsic limitations of FP-type algorithms: necessity to compute a best response at every iteration and decaying learning rate.


In this paper, we propose an algorithm to tackle both aspects: it is designed to learn master policies, and it is more efficient than FP. The algorithm is a deep RL extension of OMD to MFG with population-dependent policies. In Section~\ref{sec:background} we introduce the main elements of background. In Section~\ref{sec:algo}, we present the algorithm and explain how we overcome the main technical challenges. In Section~\ref{sec:expe}, we present numerical experiments on popular examples from the MFG literature, including comparisons of our method with four baselines. Last, we discuss our findings in Section~\ref{sec:discu} and conclude in Section~\ref{sec:conclusion}. 

\section{Background}
\label{sec:background}
In this paper, we consider a multi-agent decision-making problem using the framework of discrete-time MFGs, which relies on the classical notion of Markov decision processes. We denote by $\Delta_E$ the simplex of probability distributions on a finite set $E$.

\paragraph{MDP for a representative agent.} Specifically, we focus on a finite state space $\mathcal{X}$ and a finite action space $\mathcal{A}$. 
We denote the time horizon by $N_T$, and at each time step $n \in [0, \ldots, N_T]$, we represent the mean field term (i.e., the population's state distribution) as $\mu_n \in \Delta_{\mathcal{X}}$. When the representative agent is at state $x_n$ and takes action $a_n$, transition probability for its next state is denoted by $p(\cdot\mid x_n, a_n, \mu_n)$, and the reward it receives is $r(x_n, a_n, \mu_n)$. For simplicity, we focus on population-independent transitions, although the algorithm we propose can be directly applied to MFGs population-dependent transitions. To avoid any misunderstanding, we use bold type $\boldsymbol{\mu}$ to represent the mean field sequence, which consists of a sequence of mean field states $\mu_n$ at corresponding time steps. Given a mean field flow, the goal of the representative agent is to maximize the sum of rewards up to terminal time.  In contrast with most of the MFG literature, here the initial $\mu_0$ is not fixed a priori. This motivates the class of policies that we discuss next. 

\paragraph{Classes of policies.} Contrary to the assumption made in stationary MFG (see e.g. \citep{almulla2017two,subramanian2019reinforcement,lauriere2022learning}), which assumes a stationary population distribution over time, our focus in this paper is on a more general setting that considers an evolving process. In this kind of setting, not only does the agent's state evolve, but the population distribution also changes over time. In such cases, the agents' transitions and rewards depend on the current mean field distribution. In the finite horizon cases, the less time is left, the more urgent it becomes for agents to take action \citep{pardo2018time}. For this reason, we consider non-stationary policies. In most of the MFG literature, one considers policies $\boldsymbol{\pi} = (\pi_n)_{n=0,\dots,N_T-1}$ such that for every $n$, $\pi_n: \mathcal{X} \to \Delta_{\mathcal{A}}$. This means that the agent can look at its own state $x_n$ and choose a distribution $\pi_n(\cdot|x_n)$ over actions, from which it can pick action $a_n$. Such \emph{population-independent} policies are meaningful when the initial distribution $\mu_0$ is known and fixed in advance, because in such cases, the dependence on $\mu_n$ is implicit, through time. However, when the initial distribution is not known in advance, the agent's decision should depend on the current distribution as well. So in this work, we are interested in a more general class of policies, which have been called \emph{population-dependent} or \emph{master policies} in~\cite{Perrin2022General}. The latter work was dealing only with stationary problems but we are concerned with finite-horizon MFGs so we consider policies $\boldsymbol{\pi} = (\pi_n)_{n=0,\dots,N_T-1}$ such that for every $n$, $\pi_n: \mathcal{X} \times \Delta_{\mathcal{X}} \to \Delta_{\mathcal{A}}$. When at state $x_n$ and the population distribution is $\mu_n$, the agent picks an action $a_n$ according to the distribution $\pi_n(\cdot|x_n,\mu_n)$.


\paragraph{Best Response} 
From the perspective of a single agent, if the mean field sequence $\boldsymbol{\mu} = (\mu_n)_{n=0,\dots,N_T}$ is given, the total reward function to maximize is defined as:
\begin{equation*}
    J(\boldsymbol{\pi} ; \boldsymbol{\mu})=\mathbb{E}_{\pi}\left[\sum_{n=0}^{N_T} r_n\left(x_n, a_n, {\mu}_n\right)\right],
\end{equation*}
subject to the dynamics $ x_0 \sim \mu_0$, $x_{n+1} \sim p_n\left(\cdot \mid x_n, a_n, \mu_n\right)$, $a_n \sim \pi_n\left(\cdot \mid x_n, \mu_n\right)$, for $n = 0, \dots, N_T-1$. 
A policy $\boldsymbol{\pi}$ is a \emph{best response} to an evolutive mean field sequence $\boldsymbol{\mu}$ of the population if it maximizes the reward, namely: 
\begin{equation*}
    \boldsymbol{\pi} \in BR(\boldsymbol{\mu}):=\underset{\boldsymbol{\pi}}{\operatorname{argmax}} J(\boldsymbol{\pi} ; \boldsymbol{\mu}). 
\label{eq:br_definition}
\end{equation*}

\paragraph{Nash equilibrium. } 
We say that a pair $(\boldsymbol{\pi}, \boldsymbol{\mu})$ is a \emph{(mean field) Nash equilibrium} if $\boldsymbol{\pi}$ is a best response to $\boldsymbol{\mu}$, and $\boldsymbol{\mu}$ is generated by $\boldsymbol{\pi}$ in the sense that it is the distribution sequence generated when all the players use the policy $\boldsymbol{\pi}$. Mathematically, it means that: for $n = 0, \dots, N_T-1$, 
\looseness=-1
\begin{equation}
\label{eq:MF-induced-pi}
    \mu_{n+1}(x') = \sum_{x,a} \mu_n(x)\pi_n(a|x,\mu_n) p_n\left(x' \mid x, a, \mu_n\right).
\end{equation}
Denoting by $\boldsymbol{\mu}^{\boldsymbol{\mu}}$ the mean field sequence generated by $\boldsymbol{\pi}$, solving an MFG means finding a policy $\boldsymbol{\pi}$ such that $\boldsymbol{\pi} \in BR(\boldsymbol{\mu}^{\boldsymbol{\mu}})$, which is a fixed point problem. 
 If a population-dependent policy is an equilibrium policy for any initial distribution $\mu_0$, then we call it a \emph{master policy}. With such a policy, the game can start from any distribution and the players will always be able to play according to a Nash equilibrium without needing to learn a new policy.  


\paragraph{Exploitability}
Another way to characterize Nash equilibrium policies is through the notion of exploitability, which is a widely used metric for evaluating convergence and measuring how far a policy is from being a Nash equilibrium. Formally, it quantifies to what extent a single player can be better off by deviating from the population's behavior and using a different policy. It is defined as:
\begin{equation}
\label{exploitabiliy}
    \mathcal{E}(\boldsymbol{\pi}) = 
    \sup_{\boldsymbol{\pi}^{\prime}} J\left(\boldsymbol{\pi}^{\prime} ; \boldsymbol{\mu}^{\boldsymbol{\pi}}\right)-J_{}\left(\boldsymbol{\pi} ; \boldsymbol{\mu}^{\boldsymbol{\pi}}\right).
\end{equation}
A policy $\boldsymbol{\pi}$ is a Nash equilibrium policy if and only if $\mathcal{E}(\boldsymbol{\pi}) = 0$. When $\boldsymbol{\pi}$ is not a Nash equilibrium, its exploitability is positive, and the value of $\mathcal{E}(\boldsymbol{\pi})$ can be used to measure how far $\boldsymbol{\pi}$ is from being an equilibrium. 

Several studies have utilized the concept of exploitability, such as \citep{lockhart2019computing}. It has also been instrumental to prove the convergence of Fictitious Play~\cite{perrin2020fictitious}. Since it is not always possible to compute exact exploitability, a notion of approximate exploitability has been proposed to balance convergence measure and computational costs by using only previously computed policies rather than all policies to compute the supremum in~\eqref{exploitabiliy}. 
However, while approximate exploitability indicates learning progress, it does not capture the proximity to the genuine best response, especially in comparison with different baselines. To facilitate comparisons among different algorithms in this research, we employ the exact exploitability equation~\eqref{exploitabiliy} to gauge the deviation from the Nash equilibrium. 

\section{Algorithm}
\label{sec:algo}
In this section, we explain the main difficulties and present the algorithm we propose. Algo.\ref{algo: algorithm1} is the short version, the detailed version can be seen in Appendix Algo.\ref{algo: algorithm2}.

\subsection{Online Mirror Descent}
In the context of MFGs, Online Mirror Descent (OMD) based solution \citep{perolat2021scaling, lauriere2022scalable} is a method analogous to a kind of policy iteration with a sum over Q-functions. In the single-agent case, it corresponds to the Mirror Descent MPI algorithm, see~\cite{vieillard2020leverage}. At iteration $k$, we start by evaluating the policy $\boldsymbol\pi^{k-1}$ computed at the previous iteration, which amounts to computing its Q-function $Q^k$ that evaluates how goodness of each state-action pair as defined in classic RL, given the mean-field $\boldsymbol{\mu}^{k-1}$ induced when whole population also uses $\boldsymbol\pi^{k-1}$. Then, we aggregate this Q-function with the previous Q-functions, and we compute the new policy by taking a softmax, namely: $\pi^k_n(\cdot|x,\mu) = \operatorname{softmax}\left(\frac{1}{\tau} \sum_{i=0}^k Q^i_n(x,\mu,\cdot)\right)$. Here, for each $i$, $(Q^i_n)_{n=0,1,\dots,N_T}$ is computed by backward induction using the Bellman equation for the Q-function of the representative player facing mean field $\boldsymbol{\mu}$: 
\[
\begin{cases}
     Q^i_{N_T}(x,\mu,a) = r_{N_T}(x,\mu,a),
    \\
    Q^i_{n}(x,\mu,a) = r_{n}(x,\mu,a) 
    \\
    \,\,+ \sum_{x'} p_n(x'|x,a,\mu) \max_{a'} Q^i_{n+1}(x',\mu',a'), \, n < N_T,
\end{cases} 
\]
where the next distribution $\mu'$ is given as follows: for each $x'$, 
$
\mu'(x') = \sum_{x,a}p_n(x'|x,a,\mu)\pi_n^{i}(a|x,\mu)\mu_{n}(x)
$
Then, we compute the new mean-field flow $\boldsymbol{\mu}^{k}$ induced when the whole population also uses $\boldsymbol\pi^{k}$. The evaluation phase requires computing the mean field sequence induced by the policy using forward induction~\eqref{eq:MF-induced-pi}, but it does not require computing a best response. See Algo. \ref{algorithm3} in the Appendix.

However, since $\mu \in\Delta_{\mathcal{X}}$ takes a continuum of values, it is not possible to represent exactly $Q^k$. It also raises the question of how to represent the population distribution $\mu$ which will be discussed in Section \ref{sec:expe}. For this reason, we will resort to function approximation and use deep neural networks to approximate Q-functions. Classically, learning the Q-function associated with a policy can be done using Monte Carlo samples. Nevertheless, computing the sum of neural network Q-functions would be challenging because neural networks are non-linear approximators. To tackle this challenge, as explained below, we will use the Munchausen trick, introduced in~\cite{vieillard2020munchausen} and adapted to the MFG setting in~\cite{lauriere2022scalable}. Although the latter reference also uses an OMD-type algorithm, it does not take handle population-dependent policies and Q-functions, which is a major difficulty addressed here. This requires in particular a suitable use of a set of initial distributions and of the replay buffer. 

\subsection{Q-function update} 
A naive implementation would consist in keeping copies of past neural networks $(Q^i)_{i=0,\dots,k}$, evaluating them and summing the outputs but this would be extremely inefficient. Instead, we define a regularized Q-function and establish Thm~\ref{thm:equivalence} using the idea of Munchausen trick~\cite{vieillard2020munchausen}. It relies on computing a single Q-function that mimics the sum $\sum_{i=0}^{k-1} Q^i$ by using implicit regularization thanks to a Kullback-Leibler (KL) divergence between the new policy and the previous one. We derive, in our mean-field context, the rigorous equivalence between regularized Q and the summation of historical Q values. 

\begin{theorem}
\label{thm:equivalence}
Denote by $\boldsymbol{\pi}^{k-1}$ the softmax policy learned in iteration $k-1$, i.e., $\pi^{k-1}_n(\cdot|x,\mu) = \operatorname{softmax}\big(\frac{1}{\tau} \sum_{i=0}^{k-1} Q^i_n(x,\mu,\cdot)\big)$, and by $Q^k$  the state-action value function in iteration $k$. Let $\widetilde{Q}^k=Q^k+\tau \ln \boldsymbol{\pi}^{k-1}$, which is a function from $\mathbb{N} \times \mathcal{X} \times \Delta_{\mathcal{X}} \times \mathcal{A}$ to $\mathbb{R}$. If $\mu^k$ is generated by $\boldsymbol{\pi}^{k-1}$ in Algo.~\ref{algo: algorithm1}, then for every $n,x$, 
\begin{equation}
\label{the1}
        \pi^{k}_n(\cdot|x,{\mu_n^{k}})
        = \operatorname{softmax}\left(\frac{1}{\tau} \widetilde{Q}^k_n(x,{\mu_n^{k}},\cdot)\right).
\end{equation}
\end{theorem}

The proof relies on the following two equalities. First,
$
     \operatorname{softmax}\big(\frac{1}{\tau}\sum_{i=0}^k Q^i\big)=\operatorname{argmax}_\pi\left\langle\pi, Q^k\right\rangle-\tau \mathrm{KL}\left(\pi \| \pi^{k-1}\right),
$
and second,
$
\operatorname{softmax}\big(\frac{1}{\tau} \tilde{Q}^k\big) = \operatorname{argmax}_\pi\big\langle\pi, \tilde{Q}^k\big\rangle-\tau\langle\pi , \ln \pi\rangle $. 
We start by showing the two equalities hold, then prove the equivalence between the right-hand side of both equations. The proof process utilizes the Lagrange multiplier method and the constraint that $\boldsymbol{1}^{\top} \pi=1$. 
The detailed proof is shown in Appendix \ref{proof:softmax}. 
\looseness=-1

Based on Theorem \ref{thm:equivalence}, at iteration $k$ of our algorithm (see Algo.~\ref{algo: algorithm1}) we train a deep Q-network $\tilde{Q}^k_\theta$ with parameters $\theta$  which takes as inputs the time step, the agent's state, the mean-field state, and the agent's action. This neural network is trained to minimize the loss:
$ \mathbb{E}\left|\tilde{Q}^k_\theta\left(\left(n, x_n,\mu_n\right), a_n\right) - T_n\right|^2
$
where the target $T_n$ is: 
\begin{equation}
\begin{aligned}
    T_n &=   r_n^k +  {\color{blue}L^{k}_n} \\
    & + \gamma \sum_{a_{n+1}} \pi^{k}_{\theta^{\prime}}\left(a_{n+1} \mid s_{n+1}^k\right)\Big[
    \tilde{Q}^k_{\theta^{\prime}}\left(s_{n+1}^k, a_{n+1}\right) {\color{blue}-\, L^{k}_{n+1}}\Big]
\end{aligned}
\label{eq:target-Tn}
\end{equation}
where $r_n^i = r(x_n,a_n,\mu_n^k)$, $s_n^k = (n, x_{n}, \mu_{n}^k)$, and ${\color{blue}L^{k}_{n} = \tau \log \left(\pi^{k-1}_{\theta}\left(a_n \mid s_n^k \right)\right)}$, which is the main difference with a classical DQN target.  
Here $\pi^{k}_{\theta^{\prime}} = \operatorname{softmax}(\frac{1}{\tau} \tilde{Q}^k_{\theta^{\prime}})$ where $\tilde{Q}^k_{\theta^{\prime}}$ is a target network with same architecture but parameters $\theta^{\prime}$, updated at a slower rate than $\tilde{Q}^k_{\theta}$. In the definition of $T_n$, the {\color{blue}terms in blue} involve $\pi^{k-1}_{\theta} = \operatorname{softmax}\left(\frac{1}{\tau} \widetilde{Q}^{k-1}_{\theta}\right)$, which performs implicit averaging (see Appendix~\ref{proof:thm} for more details).

Differing from the Q update function proposed in~\citep{lauriere2022scalable}, where the target policy is set as the policy learned from the previous iteration, our algorithm employs the target policy $\pi^{k}_{\theta^{\prime}}$ and target Q-function $\tilde{Q}^k_{\theta^{\prime}}$ being learned in the current iteration, namely $k$ instead of $k-1$. This modification holds significant importance because if the target policy remains fixed by a separate policy, the distribution of the policy under evaluation would differ from the one being learned. Consequently, this distribution shift would induce extra instability in the learning process.

\begin{algorithm}
\caption{Master Deep Online Mirror Descent (short) }\label{algo: algorithm1}
\SetAlgoLined
\SetKwData{Left}{left}\SetKwData{This}{this}\SetKwData{Up}{up}
\SetKwFunction{Union}{Union}\SetKwFunction{FindCompress}{FindCompress}
\SetKwInOut{Input}{input}\SetKwInOut{Output}{output}

Initial distributions $\mathcal{D}$; Munchausen parameter $\tau$, Q-network parameter $\theta$;  target network parameter $\theta^{\prime}$,


\begin{flushleft}

Set initial 
$\boldsymbol{\pi}^0(\cdot \mid n, x,\mu)=\operatorname{softmax}\left(\frac{1}{\tau} {Q}_{\theta}(n, x,\mu, \cdot)\right)$
\end{flushleft}
\begin{flushleft}
\For{iteration $k=1,2,\dots,K$} 
{

1. Compute mean-field sequence $(\boldsymbol{\mu}^{k,\mu_0}_n)_{n=0,\dots,N_T}$ with  $\boldsymbol{\pi}^{k-1}$ for all in $\mu_0 \in \mathcal{D}$; see Eq.~\eqref{eq:MF-induced-pi}


2. Reset replay buffer $\mathcal{M_{RL}}$ 

3. Value function update with DRL:

\For{episode $t=1,2,\dots,N_{episodes}$}
 {
 \For{  ${\boldsymbol{\mu}^k}$ in $(\boldsymbol{\mu}^{k,\mu_0})_{\mu_0 \in \mathcal{D}}$}
 {
     \For{time step $n=1,2,\dots,N_T$}
        {

            Select action $a \sim \epsilon-$greedy $\tilde{Q}_{\theta}$ \\
                Execute $a_n$; Store transition in $\mathcal{M_{RL}}$\\
                Update $\theta$ (and $\theta^{\prime}$ periodically) with Eq.~\eqref{eq:target-Tn}
 
            
            
    
            
            
        }
 }
 
 }
    
4. Policy update: 
$\boldsymbol{\pi}^k(\cdot \mid n, x,\mu)=\operatorname{softmax}\left(\frac{1}{\tau} \tilde{Q}_{\theta}(n, x,\mu, \cdot)\right)$
} 
Return $\boldsymbol{\pi}^K$
\end{flushleft}

\end{algorithm}

\subsection{Inner loop replay buffer}
In order to effectively learn the master policy capable of dealing with any initial distribution, incorporating the population distribution at each time step as input and training naively the Q-function for various initial distributions is insufficient in itself. This paper makes use of a replay buffer in a specific way 
to ensure the attainment of the master policy. Experience replay buffer is a widely employed technique in RL to enhance training stability and sample efficiency \citep{mnih2015human, schaul2015prioritized}. The core concept involves storing past experiences $(x_n, a_n, r_n, \mu_n, x_{n+1})$ in a buffer and sampling a batch of experiences from this buffer during Q network training, based on certain predefined rules.

In the MFG setting we consider, as both the distribution and timestep are integral components of the states, the utilization of the replay buffer in our algorithm aligns precisely with the stationary sampling data requirements, which is consistent with the original proposal \citep{mnih2015human}. However, maintaining the buffer for the entire history would impose a substantial burden and significantly slow down the learning process, as it may repeatedly learn well-explored tuples from the past. Additionally, if we desire the neural network to learn from multiple initial distributions, we need to input multiple evolutionary training processes into the network separately. Yet, due to the phenomenon of catastrophic forgetting \citep{goodfellow2013empirical, kirkpatrick2017overcoming} in neural networks, if the gap between two evolutionary processes is too wide, the previous learning may be forgotten.

Therefore, the crucial matter lies in determining the optimal location for constructing and resetting the replay buffer. With the implicit summation of historical Q values, it is no longer necessary to sample from previous iterations during subsequent training sessions. Consequently, we choose to reset the replay buffer at the start of each iteration. In order to prevent catastrophic forgetting, we maintain the buffer and store data from various evolutive mean field sequences during each iteration. The efficacy of this placement strategy, as illustrated in Step 2 of Algo.~\ref{algo: algorithm1}, will be empirically demonstrated through experiments (see Fig.~\ref{buffer_replay} below).

\section{Experiments setup}
\label{sec:expe}
 \textbf{Environments. } The four environments chosen in this paper are canonical examples that are widely used as MFG domains. We recall that within the realm of MFGs, solving games necessitates identifying an equilibrium, which is harder than simply maximizing a reward. Furthermore, the mean field evolution depends on the initial distribution. In each experiment, we explore two different scenarios respectively. The first scenario, referred to as {\bf fixed $\mu_0$} in the sequel, follows the common practice, where the population always starts from a fixed initial distribution. The second scenario, referred to as {\bf multiple $\mu_0$} in the sequel, aims to examine the effectiveness of the master policy. In this scenario, we set different initial distributions which are simultaneously used for training. The detailed distributions can be seen in the Appendix. Instead of training multiple Nash equilibria with different networks, the master policy aims to use one single network to learn the equilibrium policies for different initial distributions. Intuitively, population-independent policies cannot perform well in this scenario (unless the equilibrium policy does not vary when the initial distribution changes, which amounts to say that there are no interactions).

\textbf{Algorithms. } Throughout the experiments, we compare our algorithm with 4 baselines, including several SOTA algorithms in the domain of Deep RL for MFGs. In the figures and tables presented in the sequel, {\bf vanilla FP (V-FP)} refers to an adaptation of (tabular) FP from \citep{perrin2020fictitious} to deep neural networks. V-FP uses classic fictitious play (see Algo. \ref{algorithm2}) to iteratively learn the Nash equilibrium, implicitly assuming agents always start from a fixed distribution. {\bf Master FP (M-FP)} is the population-dependent FP from \citep{Perrin2022General}, which aimed to handle any initial distribution via FP. {\bf Vanilla OMD1 (V-OMD1)} is the Deep OMD introduced in \citep{lauriere2022scalable} based on Munchausen trick. {\bf Vanilla OMD2 (V-OMD2)} is our algorithm \emph{without} the input of mean field state, while our full algorithm is caleld {\bf Master OMD (M-OMD)}. With this terminology, M-FP and M-OMD learn population-dependent policies, while V-FP, V-OMD1 and V-OMD2 do not. In particular, V-OMD2 can be viewed as an ablation study of our main algorithm (M-OMD), where we remove the distribution dependence to see the performance. 

{\bf Implementation. } In FP-type algorithms, we utilize the DQN algorithm to learn the best response with respect to the current mean-field sequence and iterate this process multiple times. Notably, in the model-free setting, the transition probability and reward function are unknown during both training and execution. The structure of the Q network in this paper follows that of the DQN \citep{mnih2015human}, \citep{ota2021training}. Likewise, in OMD-type algorithms, policy evaluation is done using a Q-network with similar architecture. 
The distribution is represented as a histogram. We used a one-dimensional vector (in which all the dimensions are concatenated if needed) and then passed it to the neural network. Our experiments show that, in the examples we study, this approach works well. A possible improvement, particularly for 2D examples, would be to use ConvNets as in~\cite{Perrin2022General}, but this was not needed in our experiments. Importantly, our method learns \emph{non-stationary} policies. Indeed, this is necessary for finite-horizon problems: in contrast with the infinite horizon stationary setting (see e.g.~\cite{Perrin2022General}), timesteps are here an essential information that should be incorporated into the agent's policy. To effectively perceive the timestep, we employ a one-hot encoding to convert the scalar timestep into a one-hot vector. Last, for each algorithm, we swept over hyperparameters and used the best parameters we found.

The GPU used is NVIDIA TITAN RTX (24gb), the CPU is 2x 16-core Intel Xeon Gold (64 virtual cores). The exploitability curves are averaged over 5 realizations of the algorithm, and whenever relevant, we show the standard deviation (std dev) with a shaded area. The details about training and hyperparameters are listed in the Appendix.

\subsection{Exploration}
\label{experiment:exploration}
Exploration is a classic problem in MFG \citep{geist2021concave}, in which the a large group of agents tries to uniformly distribute into empty areas but in a decentralized way. In this section, we introduce two variants, with different geometries of domain. The first is in a big empty room, and the second is in four connected rooms, which makes the problem much more challenging.

\paragraph{Example 1: Exploration in one room. }
We consider a 2D  grid world of dimension $11 \times 11$. The action set is $\mathcal{A}$=\{up, down, left, right, stay\}. The dynamics are:
$    x_{n+1}=x_n+a_n+\epsilon_n,$ 
where $\epsilon_n$ is an environment noise that perturbs each agent's movement (no perturbation w.p. $0.9$, and one of the four directions w.p. $0.025$ for each direction). 
The reward function will discourage agents from being in a crowded location:
 $   
 r(x, a, \mu)=-\log (\mu(x))-\frac{1}{|X|}|a|.
 $

\paragraph{Example 2: Exploration in four connected rooms. }

The same reward function is the same and the dynamics is similar but the environment is four connected rooms and the agent does not move when it hits an obstacle. The goal is to explore every grid point in those rooms. The dimension of the whole map is also $11\times11$ (a comparative study with different map sizes is shown in Fig.~\ref{fig:map_size}).

\begin{figure}[htb]
\centering
\subfloat[Evolution process]{
   \includegraphics[width=0.45\columnwidth]{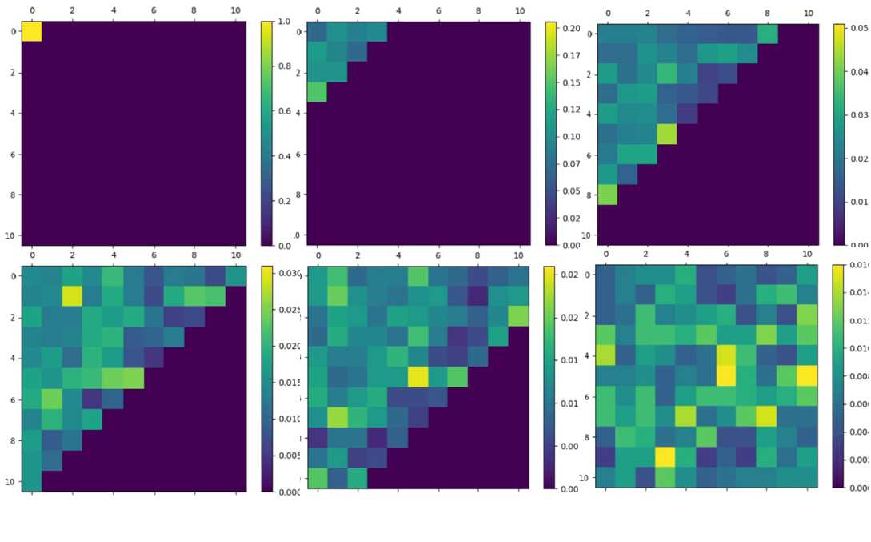}
}
\subfloat[Exploitability (fixed $\mu_0$)]{
   \includegraphics[width=0.45\columnwidth]{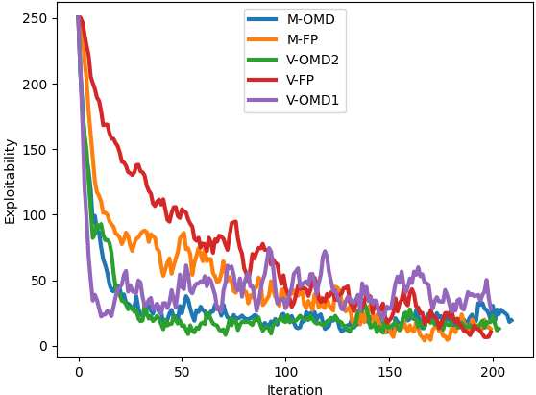}
}\\ 
\subfloat[Exploitability (multiple $\mu_0$)\,]{
   \includegraphics[width=0.45\columnwidth]{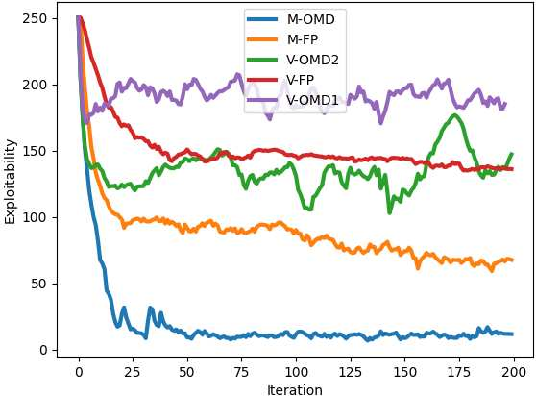}
}\,\,
\subfloat[Exploitability (multiple $\mu_0$)]{
   \includegraphics[width=0.45\columnwidth]{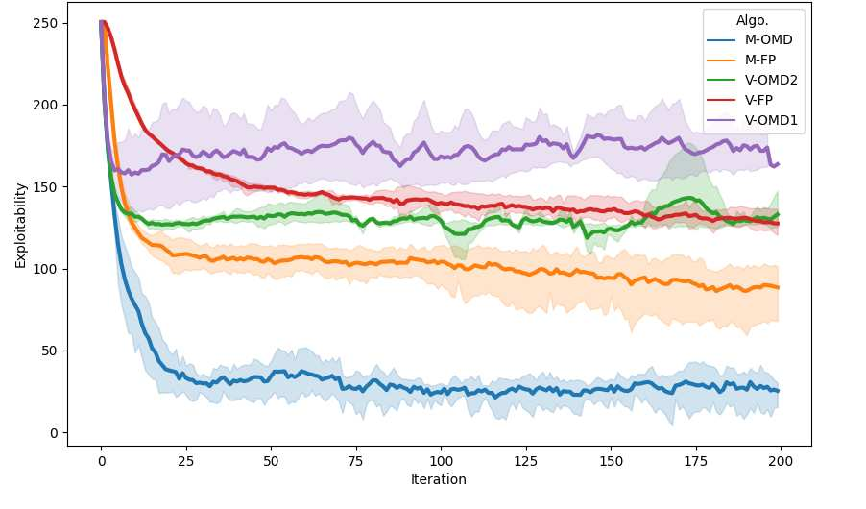}
}
\caption{Example 1: Exploration in one room. (a): density evolution using the policy learnt by M-OMD, starting from the $\mu_0$ used for (b). (b): exploitability vs training iteration for a single $\mu_0$. (c): average exploitability when training over 5 different $\mu_0$ (single run of each algo.). (d): averaged curve over 5 runs and std dev.}
\label{uniform_fig}
\end{figure}

\begin{figure}[htb]
\centering
\subfloat[Evolution process]{
\begin{minipage}[t]{0.5\linewidth}
\centering
\includegraphics[width=1.5in]{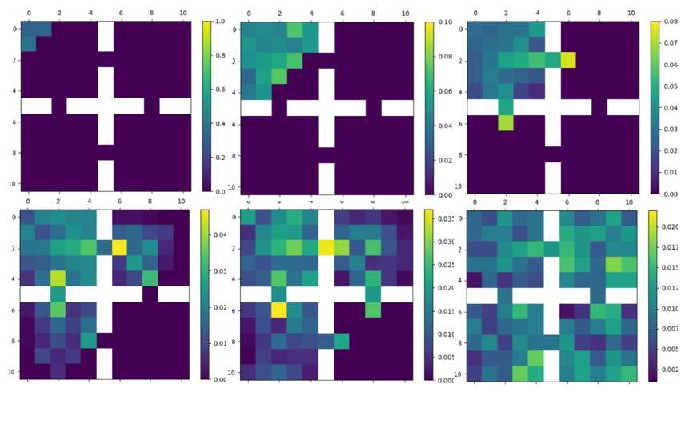}
\end{minipage}%
}%
\subfloat[Exploitability (fixed $\mu_0$)]{
\begin{minipage}[t]{0.5\linewidth}
\centering
\includegraphics[width=1.45in]{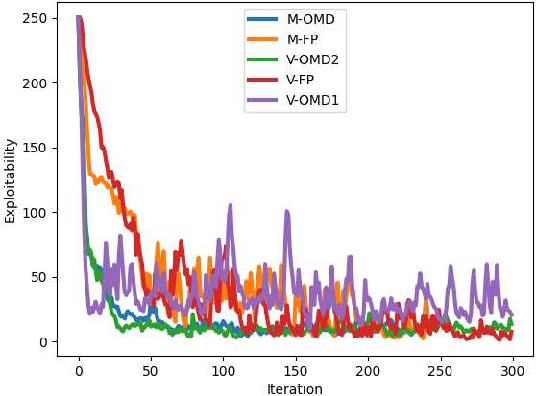}
\end{minipage}%
}%
\newline
\subfloat[Exploitability (multiple $\mu_0$)]{
\begin{minipage}[t]{0.55\linewidth}
\centering
\includegraphics[width=1.45in]{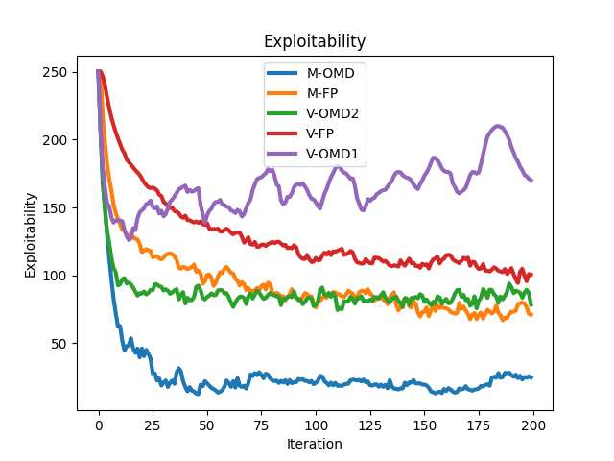}
\end{minipage}%
}%
\subfloat[Exploitability (multiple $\mu_0$)]{
\begin{minipage}[t]{0.45\linewidth}
\centering
\includegraphics[width=1.65in]{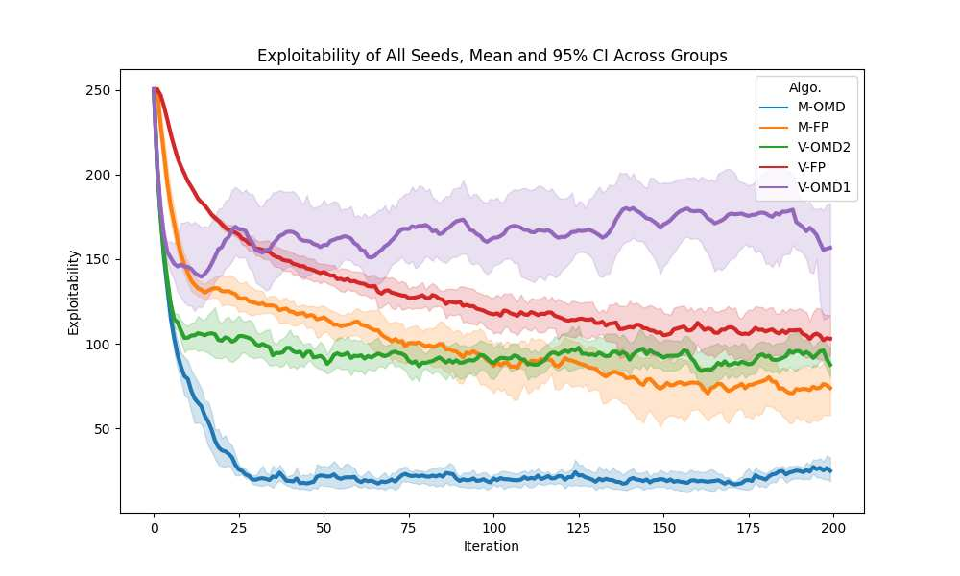}
\end{minipage}
}%
\centering
\caption{Example 2: Exploration in four connected rooms. 
(a): density evolution using the policy learnt by M-OMD, starting from the $\mu_0$ used for (b). (b): exploitability vs training iteration for a single $\mu_0$. (c): average exploitability when training over 5 different $\mu_0$ (single run of each algo.). (d): average over 5 runs \& std dev.
}
\label{exploration_fig}
\end{figure}

Fig.~\ref{uniform_fig} and Fig.~\ref{exploration_fig} show the results for Ex.~1 and Ex.~2 respectively: heat-maps representing the evolution of the distribution (at several time steps), when using the master policy learnt by our algorithm; evolution exploitability when using a fixed $\mu_0$ or multiple $\mu_0$ for a single run of the algorithm; and finally the results averaged over 5 runs. On both examples, our proposed algorithm (M-OMD) converges faster than all the 4 baselines. With fixed $\mu_0$, all methods perform well, but with multiple initial distributions, it appears clearly that F-FP, V-OMD1 and V-OMD2 fail to converge, which is due to the fact that vanilla policies lack awareness of the population so agents cannot adjust their behavior suitably when the initial distribution varies. In other words, population-independent policies cannot be Nash equilibrium policies when testing on new initial distributions, and hence their exploitability is non-zero.

\begin{figure}[htbp]
  \centering
    \begin{minipage}[t]{0.8\linewidth}
    \centering
    \includegraphics[width=2.5in]{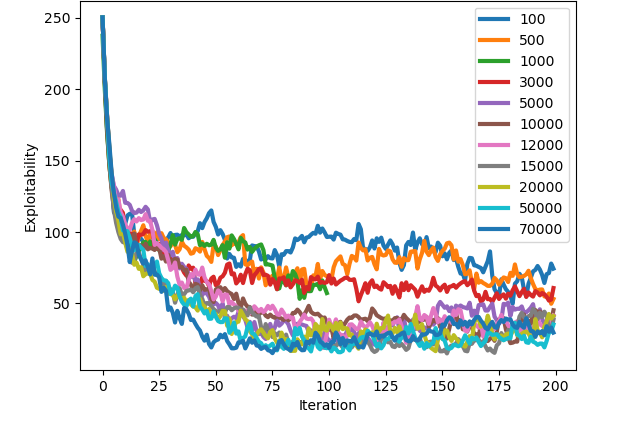}
    \end{minipage}%
  \caption{Exploitability vs iteration number for various buffer sizes, using our M-OMD algorithm, in exploration of four connected room task (see Sec.~\ref{experiment:exploration}). Small sizes lead to the forgetting of some $\mu_0$ and hence poor performance (see Step 2 in Algo.~\ref{algo: algorithm1}). } 
\label{buffer_replay}
\end{figure}

\subsection{Example 3: Beach bar}

The Beach bar environment, introduced in~\citep{perrin2020fictitious} represents agents moving on a beach towards a bar. The goal for each agent is to (as much as possible) avoid the crowd but get close to the bar. The dynamics are the same as in the exploration examples. Here we consider that the bar is located at the center of the beach, and that there are walls on the four sides of the domain. The reward function is:
$
r\left(x, a, \mu\right)={d_{bar}}\left(x\right)-\frac{\left|a\right|}{|\mathcal{X}|}-\log \left(\mu\left(x\right)\right), 
$
where $d_{bar}$ indicates the distance to the bar, the second term penalizes movement so the agent moves only if it is necessary, and the third term penalizes the fact of being in a crowded region. Here, we consider $\mathcal{X}$ with $11$ (1D) or $11\times11$ (2D) states. 

\begin{figure} [htb]
\centering
\subfloat[Evolution process in 2D]{
\begin{minipage}[t]{0.9\linewidth}
\centering
\includegraphics[width=3in]{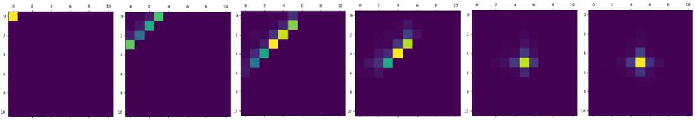}
\end{minipage}
}%
\newline
\subfloat[Evolution process in 1D]{
\begin{minipage}[t]{0.5\linewidth}
\centering
\includegraphics[width=1.6in]{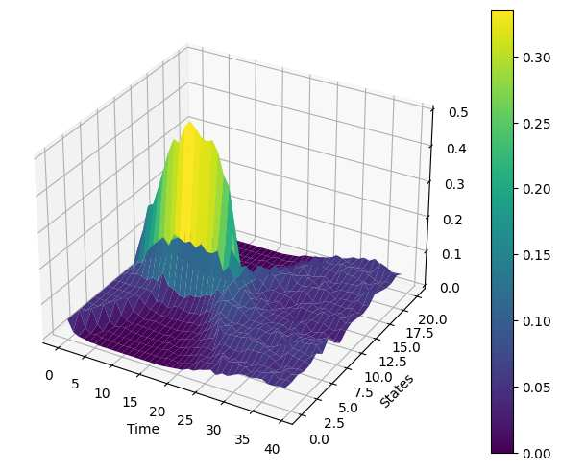}
\end{minipage}
}%
\subfloat[Exploitability (fixed $\mu_0$)]{
\begin{minipage}[t]{0.5\linewidth}
\centering
\includegraphics[width=1.45in]{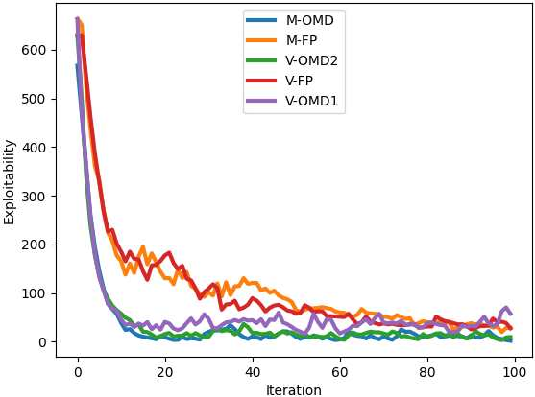}
\end{minipage}
}\
\subfloat[Exploitability (multiple $\mu_0$)]{
\begin{minipage}[t]{0.55\linewidth}
\centering
\includegraphics[width=1.45in]{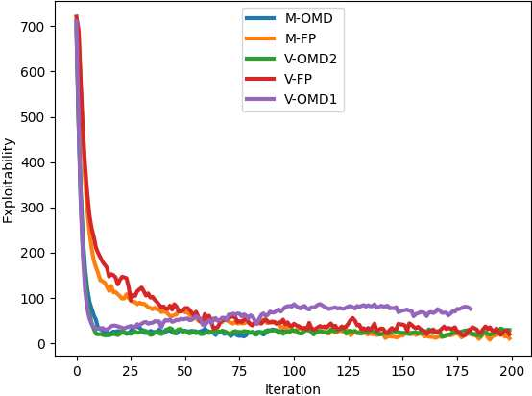}
\end{minipage}
}%
\subfloat[Exploitability (multiple $\mu_0$)]{
\begin{minipage}[t]{0.45\linewidth}
\centering
\includegraphics[width=1.55in]{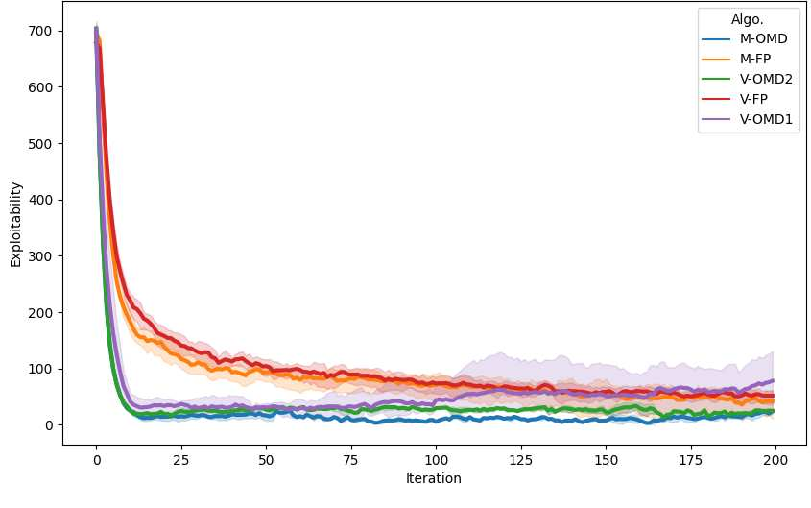}
\end{minipage}
}
\caption{Example 3: Beach bar problem. (a) and (b) show the distribution evolution for 2D and 1D case. (c), (d) and (e): exploitability for 2D case with: (c) when training with fixed $\mu_0$, (d) when training with different $\mu_0$, and (e) when training with different $\mu_0$ and averaging over 5 runs.
}
\label{bar_fig}
\end{figure}

The results are shown in Fig.~\ref{bar_fig}. The agents are always attracted towards the bar, whose location is independent of the population distribution. So the impact of starting from a fixed distribution or from various distributions is expected to be relatively minimal, which is also what we observe numerically. (a) is for a 2D domain and we see that the distribution spreads while the agents move towards the bar to avoid large crowds, and then focuses on the bar's location. (b) is for a 1D model but we add an extra difficulty: the bar closes at time $n=20$ and hence the population goes back to a uniform distribution (due to crowd aversion). This shows that the policy is aware of time. (c) and (d) show the performance when training over multiple $\mu_0$. Here again, M-OMD performs best.

\subsection{Example 4: Linear-Quadratic}
In this section, we provide results on a fourth example. It is a linear quadratic (LQ) model, which is a classical setting that has been studied extensively; see e.g. \citep{carmona2013control}, \citep{bensoussan2016linear}. A discretized version was introduced in~\citep{perrin2020fictitious}. It is a 1D model, in which the dynamics are:
$x_{n+1}=x_n+a_n \Delta_n+\sigma \epsilon_n \sqrt{\Delta_n}$, 
where $\mathcal{A}=\{-M, \ldots, M\}$, corresponding to moving by a corresponding number of states (left or right). The state space is $\mathcal{X}=\{-L, \ldots, L\}$, of dimension $|\mathcal{X}|$ =$2L-1$. To add stochasticity into this model, $\epsilon_n$ is an additional noise will perturb the action choice with $\epsilon_n \sim \mathcal{N}(0,1)$, but was discretized over $\{-3 \sigma, \ldots, 3 \sigma\}$. The reward function is:
\[
    r_n\left(x,a,\mu\right)=\big[-\tfrac{1}{2}\left|a\right|^2+q a\left(m-x\right)-\tfrac{\kappa}{2}\left(m-x\right)^2\big] \Delta_n
\]
where $m=\sum_{x \in \mathcal{X}} x \mu(x)$ is the first moment of population distribution which serves as the reward to encourage agents to move to the population's average but also tries to keep dynamic movement. The terminal reward is $r_{N_T}\left(x, a, \mu\right)=-\frac{c_{\text {term }}}{2}\left(m-x\right)^2$. Here we used $\sigma=1$, $N_T=30$, $\Delta_n=1$, $q=0.01$, $\kappa=0.5$, $K=1$ $M=3$, $L=20$, $c_{term}=1$. 

\begin{figure}[htb]
\centering
\subfloat[Evolution process]{
\includegraphics[width=0.45\linewidth]{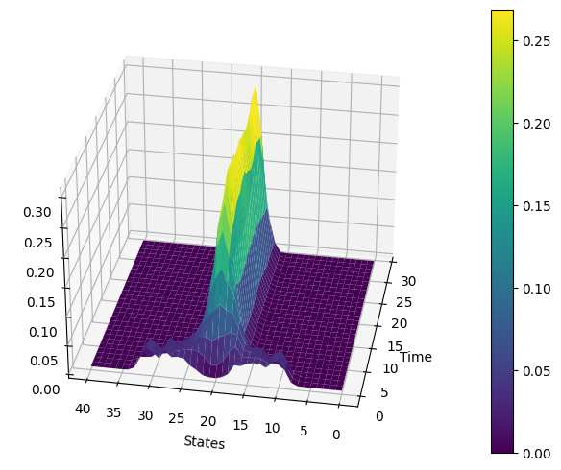}
}%
\newline
\subfloat[Exploitability (fixed $\mu_0$)]{
\includegraphics[width=0.4\linewidth]{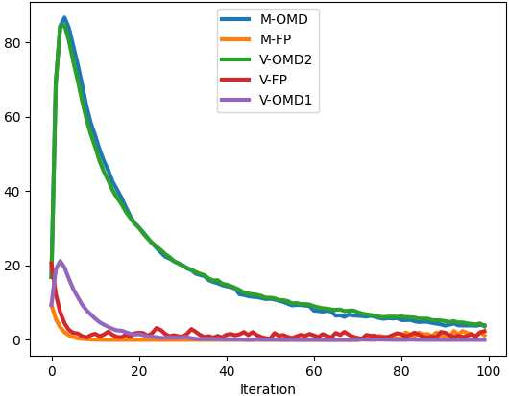}
}\,\,%
\subfloat[Exploitability (multiple $\mu_0$)]{
\includegraphics[width=0.45\linewidth]{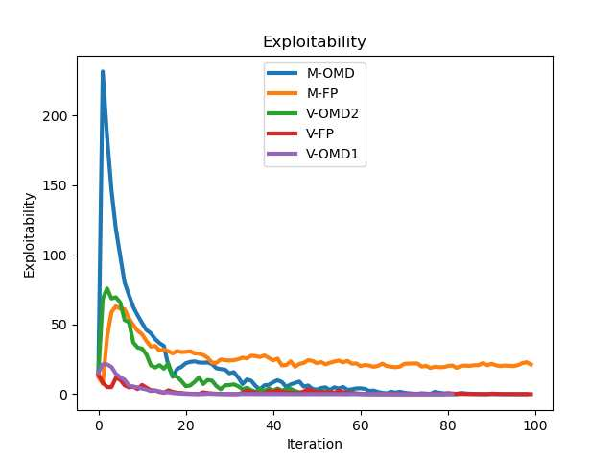}
}%
\caption{Example 4: Linear quadratic model. (a) shows the evolution of the population using the policy learned by the M-OMD algorithm, starting from two Gaussian distribution pairs, and then accumulating into the center of the population. (b) and (c) shows the averaged exploitability obtained during training over one fixed initial distribution and five initial distributions, respectively.} 
\label{lq_fig}
\end{figure}

Compared with the tasks mentioned above, solving this LQ game is considerably easier and convergence occurs with only a few iterations. However, this gives rise to some counter-intuitive phenomena in the numerical results. Our algorithm rely on modified Bellman equations, which incorporate an additional term as a regularizer to prevent a rapid change of policy during training. Consequently, our algorithm for LQ does not converge to the Nash equilibrium as fast as FP. This desirable degeneration is shown in Fig.~\ref{lq_fig}, where FP and V-OMD1 demonstrate faster convergence compared to V-OMD2 and M-OMD. In the case of V-OMD1, the parameters we used are small regularized coefficients than M-OMD and V-OMD2 (see sweeping results in Fig.~\ref{fig: sweep} in Appendix) resulting in faster convergence than our algorithms, though still slower than FP. Regarding the exploitability in multiple initial distribution training, in theory, the population should gravitate towards the center of the whole population. However, the results indicate that even vanilla FP or OMDs can decrease to zero. Our analysis is that since the moving cost is cheaper than the rewards agents receive, vanilla policies can learn a strategy that moves all agents to a specific position regardless of the initial distributions. Our tests also revealed that if the cost of moving is too high, all algorithms learn a policy that keeps agents stationary. Finding more appropriate values for the LQ model's parameters to demonstrate the influence of population-dependence is deferred to future research.

We conclude this section with Table~\ref{table:exploitability_testing}, showing a comparison of the exploitabilities. M-OMD consistenly perform better than the SOTA baseline M-FP of~\cite{Perrin2022General}. It also performs better than baselines learning population-independent policies, except in the LQ-case, where we conjecture that the equilibrium policies are almost population-independent, at least in the setting studied in the literature.
\begin{table}[htb]
\caption{Exploitability in testing set after 200 training iterations}
\label{table:exploitability_testing}
\vskip 0.1in
\begin{center}
\begin{scriptsize} 
\begin{sc}
\setlength\tabcolsep{5pt} 
\begin{tabular}{lccccc}
\toprule
Env. & V-FP & M-FP & V-OMD1 & V-OMD2 & \textbf{M-OMD}\\
\midrule
Exploration-1 & 135.32 & 84.76 & 175.91 & 163.42 & \textbf{78.2} \\
Exploration-2 & 146.45 & 72.66 & 166.84 & 159.67 & \textbf{60.0} \\
Beach Bar & 83.24 & 40.12 & 80.81 & 61.71 & \textbf{22.05} \\
LQ & 0 & 22.78 & 0 & 0 & \textbf{8.20} \\
\bottomrule
\end{tabular}
\end{sc}
\end{scriptsize}
\end{center}
\vskip -0.1in
\end{table}

\section{Discussion}
\label{sec:discu}

Building on what has been observed in the numerical experiments, we provide extra comments on a heuristic comparison between the various algorithms we studied. V-OMD1 is the the Munchausen deep OMD algorithm introduced in~\citep{lauriere2022scalable}. However, V-OMD2 differs from V-OMD1 in the following key aspects. First, V-OMD1 employs the old policy learned from the previous iteration as both the behavior policy and the target policy, consistently throughout, which can hinder convergence stabilization in some of the examples. Second, instead of using an individual hyperparameter in~\eqref{eq:target-Tn}, we utilize a clip function. The stabilized parameter is designed to avoid singularity, and a value of $10^{-6}$ suffices for this purpose. 
Third, we demonstrate the advantage of incorporating an inner-loop replay buffer to prevent catastrophic forgetting, which hinders the convergence of the population-dependent policy. 

In comparison with the experiments reported in M-FP~\citep{Perrin2022General}, in 2D scenarios, our algorithm exhibits excellent performance without employing ConvNets. In addition, numerical results demonstrate that our algorithm achieves faster convergence than M-FP. This is probably due to the fact that FP-based methods have an updating rate decaying with iterations, and that each iteration typically computes a pure policy, which is not the case with OMD.
Additionally, FP-based methods typically require averaging over a growing number of past iterations, while implicit regularization with Munchausen trick allows the OMD implementation to  avoid this issue. 
Furthermore, our algorithm is also more efficient than M-FP in terms of memory, see Fig.~\ref{memory_cost} in the Appendix. 
%
%
Last, our algorithm can be applied to a variety of other examples. We illustrate this flexibility on a variant of the exploration problem, in which we introduced ad-hoc teaming (see Appendix~\ref{sec:adhoc-teaming}). This involves adding more agents to the game randomly during the middle of the game, simulating real-life scenarios. 

\section{Conclusion}
\label{sec:conclusion}
This paper presents an algorithm called Master OMD (M-OMD) for computing population-dependent Nash equilibria in MFGs, which is more efficient than the SOTA algorithm (M-FP). By extending the Munchausen OMD algorithm to population-aware functions, we propose an effective Q-updating rule that enables the realization of this algorithm. In contrast to stationary MFGs and finite horizon MFGs assuming a fixed initial distribution, our work focuses on models where the initial population is a priori unknown and evolves over time. We allow agents to learn policies that lead to a Nash equilibrium from various initial distributions. Extensive numerical experiments demonstrate clearly the advantages of our proposed M-OMD algorithm over baselines. We leave for future work the theoretical analysis, such as a proof of convergence, perhaps under monotonicity conditions. Furthermore, it would be interesting to extend the algorithm to other settings, such as multi-population MFGs. 



\clearpage
\section{Impact Statement}
This paper presents work whose goal is to advance the field of Machine Learning. There are many potential societal consequences of our work, none which we feel must be specifically highlighted here.

\section*{Acknowledgements} 
Mathieu Lauriere is affiliated with the Shanghai Frontiers Science Center of Artificial Intelligence and Deep Learning, and with the NYU-ECNU Institute of Mathematical Sciences, NYU Shanghai, 567 West Yangsi Road, Shanghai, 200126, People’s Republic of China. 

\bibliography{masteromdbib}

\begin{thebibliography}{43}
\providecommand{\natexlab}[1]{#1}
\providecommand{\url}[1]{\texttt{#1}}
\expandafter\ifx\csname urlstyle\endcsname\relax
  \providecommand{\doi}[1]{doi: #1}\else
  \providecommand{\doi}{doi: \begingroup \urlstyle{rm}\Url}\fi

\bibitem[Almulla et~al.(2017)Almulla, Ferreira, and Gomes]{almulla2017two}
N.~Almulla, R.~Ferreira, and D.~Gomes.
\newblock Two numerical approaches to stationary mean-field games.
\newblock \emph{Dynamic Games and Applications}, 7:\penalty0 657--682, 2017.

\bibitem[Anahtarci et~al.(2023)Anahtarci, Kariksiz, and Saldi]{anahtarci2023q}
B.~Anahtarci, C.~D. Kariksiz, and N.~Saldi.
\newblock Q-learning in regularized mean-field games.
\newblock \emph{Dynamic Games and Applications}, 13\penalty0 (1):\penalty0 89--117, 2023.

\bibitem[Bensoussan et~al.(2013)Bensoussan, Frehse, and Yam]{bensoussan2013mean}
A.~Bensoussan, J.~Frehse, and P.~Yam.
\newblock \emph{Mean field games and mean field type control theory}, volume 101.
\newblock Springer, 2013.

\bibitem[Bensoussan et~al.(2016)Bensoussan, Sung, Yam, and Yung]{bensoussan2016linear}
A.~Bensoussan, K.~Sung, S.~C.~P. Yam, and S.-P. Yung.
\newblock Linear-quadratic mean field games.
\newblock \emph{Journal of Optimization Theory and Applications}, 169:\penalty0 496--529, 2016.

\bibitem[Berger(2007)]{berger2007brown}
U.~Berger.
\newblock Brown's original fictitious play.
\newblock \emph{Journal of Economic Theory}, 135\penalty0 (1):\penalty0 572--578, 2007.

\bibitem[Brown(1951)]{brown1951iterative}
G.~W. Brown.
\newblock Iterative solution of games by fictitious play.
\newblock \emph{Act. Anal. Prod Allocation}, 13\penalty0 (1):\penalty0 374, 1951.

\bibitem[Burmeister et~al.(1997)Burmeister, Haddadi, and Matylis]{burmeister1997application}
B.~Burmeister, A.~Haddadi, and G.~Matylis.
\newblock Application of multi-agent systems in traffic and transportation.
\newblock \emph{IEE Proceedings-Software}, 144\penalty0 (1):\penalty0 51--60, 1997.

\bibitem[Cardaliaguet and Hadikhanloo(2017)]{cardaliaguet2017learning}
P.~Cardaliaguet and S.~Hadikhanloo.
\newblock Learning in mean field games: the fictitious play.
\newblock \emph{ESAIM: Control, Optimisation and Calculus of Variations}, 23\penalty0 (2):\penalty0 569--591, 2017.

\bibitem[Cardaliaguet et~al.(2019)Cardaliaguet, Delarue, Lasry, and Lions]{cardaliaguet2019master}
P.~Cardaliaguet, F.~Delarue, J.-M. Lasry, and P.-L. Lions.
\newblock \emph{The master equation and the convergence problem in mean field games:(ams-201)}.
\newblock Princeton University Press, 2019.

\bibitem[Carmona and Delarue(2018)]{carmona2018probabilistic}
R.~Carmona and F.~Delarue.
\newblock \emph{Probabilistic theory of mean field games with applications I-II}.
\newblock Springer, 2018.

\bibitem[Carmona et~al.(2013)Carmona, Delarue, and Lachapelle]{carmona2013control}
R.~Carmona, F.~Delarue, and A.~Lachapelle.
\newblock Control of {M}c{K}ean--{V}lasov dynamics versus mean field games.
\newblock \emph{Mathematics and Financial Economics}, 7:\penalty0 131--166, 2013.

\bibitem[Chung et~al.(2018)Chung, Paranjape, Dames, Shen, and Kumar]{chung2018survey}
S.-J. Chung, A.~A. Paranjape, P.~Dames, S.~Shen, and V.~Kumar.
\newblock A survey on aerial swarm robotics.
\newblock \emph{IEEE Transactions on Robotics}, 34\penalty0 (4):\penalty0 837--855, 2018.

\bibitem[Cucker and Smale(2007)]{cucker2007emergent}
F.~Cucker and S.~Smale.
\newblock Emergent behavior in flocks.
\newblock \emph{IEEE Transactions on automatic control}, 52\penalty0 (5):\penalty0 852--862, 2007.

\bibitem[Cui and Koeppl(2021)]{cui2021approximately}
K.~Cui and H.~Koeppl.
\newblock Approximately solving mean field games via entropy-regularized deep reinforcement learning.
\newblock In \emph{International Conference on Artificial Intelligence and Statistics}, pages 1909--1917. PMLR, 2021.

\bibitem[Dorri et~al.(2018)Dorri, Kanhere, and Jurdak]{dorri2018multi}
A.~Dorri, S.~S. Kanhere, and R.~Jurdak.
\newblock Multi-agent systems: A survey.
\newblock \emph{Ieee Access}, 6:\penalty0 28573--28593, 2018.

\bibitem[Geist et~al.(2021)Geist, P{\'e}rolat, Lauri{\`e}re, Elie, Perrin, Bachem, Munos, and Pietquin]{geist2021concave}
M.~Geist, J.~P{\'e}rolat, M.~Lauri{\`e}re, R.~Elie, S.~Perrin, O.~Bachem, R.~Munos, and O.~Pietquin.
\newblock Concave utility reinforcement learning: the mean-field game viewpoint.
\newblock \emph{arXiv preprint arXiv:2106.03787}, 2021.

\bibitem[Goodfellow et~al.(2013)Goodfellow, Mirza, Xiao, Courville, and Bengio]{goodfellow2013empirical}
I.~J. Goodfellow, M.~Mirza, D.~Xiao, A.~Courville, and Y.~Bengio.
\newblock An empirical investigation of catastrophic forgetting in gradient-based neural networks.
\newblock \emph{arXiv preprint arXiv:1312.6211}, 2013.

\bibitem[Guo et~al.(2019)Guo, Hu, Xu, and Zhang]{guo2019learning}
X.~Guo, A.~Hu, R.~Xu, and J.~Zhang.
\newblock Learning mean-field games.
\newblock \emph{Advances in Neural Information Processing Systems}, 32, 2019.

\bibitem[Hadikhanloo(2017)]{hadikhanloo2017learning}
S.~Hadikhanloo.
\newblock Learning in anonymous nonatomic games with applications to first-order mean field games.
\newblock \emph{arXiv preprint arXiv:1704.00378}, 2017.

\bibitem[Hadikhanloo(2018)]{hadikhanloo2018learning}
S.~Hadikhanloo.
\newblock \emph{Learning in mean field games}.
\newblock PhD thesis, Universit{\'e} Paris sciences et lettres, 2018.

\bibitem[Hadikhanloo and Silva(2019)]{hadikhanloo2019finite}
S.~Hadikhanloo and F.~J. Silva.
\newblock Finite mean field games: fictitious play and convergence to a first order continuous mean field game.
\newblock \emph{Journal de Math{\'e}matiques Pures et Appliqu{\'e}es}, 132:\penalty0 369--397, 2019.

\bibitem[Huang et~al.(2007)Huang, Caines, and Malham{\'e}]{huang2007large}
M.~Huang, P.~E. Caines, and R.~P. Malham{\'e}.
\newblock Large-population cost-coupled lqg problems with nonuniform agents: individual-mass behavior and decentralized $\varepsilon $-nash equilibria.
\newblock \emph{IEEE transactions on automatic control}, 52\penalty0 (9):\penalty0 1560--1571, 2007.

\bibitem[Kirkpatrick et~al.(2017)Kirkpatrick, Pascanu, Rabinowitz, Veness, Desjardins, Rusu, Milan, Quan, Ramalho, and Grabska-Barwinska]{kirkpatrick2017overcoming}
J.~Kirkpatrick, R.~Pascanu, N.~Rabinowitz, J.~Veness, G.~Desjardins, A.~A. Rusu, K.~Milan, J.~Quan, T.~Ramalho, and A.~Grabska-Barwinska.
\newblock Overcoming catastrophic forgetting in neural networks.
\newblock \emph{Proceedings of the national academy of sciences}, 114\penalty0 (13):\penalty0 3521--3526, 2017.

\bibitem[Lasry and Lions(2007)]{lasry2007mean}
J.-M. Lasry and P.-L. Lions.
\newblock Mean field games.
\newblock \emph{Japanese journal of mathematics}, 2\penalty0 (1):\penalty0 229--260, 2007.

\bibitem[Lauri{\`e}re et~al.(2022{\natexlab{a}})Lauri{\`e}re, Perrin, Geist, and Pietquin]{lauriere2022learning}
M.~Lauri{\`e}re, S.~Perrin, M.~Geist, and O.~Pietquin.
\newblock Learning mean field games: A survey.
\newblock \emph{arXiv preprint arXiv:2205.12944}, 2022{\natexlab{a}}.

\bibitem[Lauri{\`e}re et~al.(2022{\natexlab{b}})Lauri{\`e}re, Perrin, Girgin, Muller, Jain, Cabannes, Piliouras, P{\'e}rolat, {\'E}lie, and Pietquin]{lauriere2022scalable}
M.~Lauri{\`e}re, S.~Perrin, S.~Girgin, P.~Muller, A.~Jain, T.~Cabannes, G.~Piliouras, J.~P{\'e}rolat, R.~{\'E}lie, and O.~Pietquin.
\newblock Scalable deep reinforcement learning algorithms for mean field games.
\newblock In \emph{International Conference on Machine Learning}, pages 12078--12095. PMLR, 2022{\natexlab{b}}.

\bibitem[Li et~al.(2019)Li, Qin, Jiao, Yang, Wang, Wang, Wu, and Ye]{li2019efficient}
M.~Li, Z.~Qin, Y.~Jiao, Y.~Yang, J.~Wang, C.~Wang, G.~Wu, and J.~Ye.
\newblock Efficient ridesharing order dispatching with mean field multi-agent reinforcement learning.
\newblock In \emph{The world wide web conference}, pages 983--994, 2019.

\bibitem[Lockhart et~al.(2019)Lockhart, Lanctot, P{\'e}rolat, Lespiau, Morrill, Timbers, and Tuyls]{lockhart2019computing}
E.~Lockhart, M.~Lanctot, J.~P{\'e}rolat, J.-B. Lespiau, D.~Morrill, F.~Timbers, and K.~Tuyls.
\newblock Computing approximate equilibria in sequential adversarial games by exploitability descent.
\newblock In \emph{Proceedings of the 28th International Joint Conference on Artificial Intelligence}, pages 464--470, 2019.

\bibitem[Lowe et~al.(2017)Lowe, Wu, Tamar, Harb, Pieter~Abbeel, and Mordatch]{lowe2017multi}
R.~Lowe, Y.~I. Wu, A.~Tamar, J.~Harb, O.~Pieter~Abbeel, and I.~Mordatch.
\newblock Multi-agent actor-critic for mixed cooperative-competitive environments.
\newblock \emph{Advances in neural information processing systems}, 30, 2017.

\bibitem[Mnih et~al.(2015)Mnih, Kavukcuoglu, Silver, Rusu, Veness, Bellemare, Graves, Riedmiller, Fidjeland, and Ostrovski]{mnih2015human}
V.~Mnih, K.~Kavukcuoglu, D.~Silver, A.~A. Rusu, J.~Veness, M.~G. Bellemare, A.~Graves, M.~Riedmiller, A.~K. Fidjeland, and G.~Ostrovski.
\newblock Human-level control through deep reinforcement learning.
\newblock \emph{Nature}, 518\penalty0 (7540):\penalty0 529--533, 2015.

\bibitem[Olfati-Saber(2006)]{olfati2006flocking}
R.~Olfati-Saber.
\newblock Flocking for multi-agent dynamic systems: Algorithms and theory.
\newblock \emph{IEEE Transactions on automatic control}, 51\penalty0 (3):\penalty0 401--420, 2006.

\bibitem[Ota et~al.(2021)Ota, Jha, and Kanezaki]{ota2021training}
K.~Ota, D.~K. Jha, and A.~Kanezaki.
\newblock Training larger networks for deep reinforcement learning.
\newblock \emph{arXiv preprint arXiv:2102.07920}, 2021.

\bibitem[Pardo et~al.(2018)Pardo, Tavakoli, Levdik, and Kormushev]{pardo2018time}
F.~Pardo, A.~Tavakoli, V.~Levdik, and P.~Kormushev.
\newblock Time limits in reinforcement learning.
\newblock In \emph{International Conference on Machine Learning}, pages 4045--4054. PMLR, 2018.

\bibitem[Perolat et~al.(2022)Perolat, Perrin, Elie, Lauri{\`e}re, Piliouras, Geist, Tuyls, and Pietquin]{perolat2021scaling}
J.~Perolat, S.~Perrin, R.~Elie, M.~Lauri{\`e}re, G.~Piliouras, M.~Geist, K.~Tuyls, and O.~Pietquin.
\newblock Scaling mean field games by online mirror descent.
\newblock \emph{Proceedings of the 21st International Conference on Autonomous Agents and Multiagent Systems}, pages 1028--1037, 2022.

\bibitem[Perrin et~al.(2020)Perrin, P{\'e}rolat, Lauri{\`e}re, Geist, Elie, and Pietquin]{perrin2020fictitious}
S.~Perrin, J.~P{\'e}rolat, M.~Lauri{\`e}re, M.~Geist, R.~Elie, and O.~Pietquin.
\newblock Fictitious play for mean field games: Continuous time analysis and applications.
\newblock \emph{Advances in Neural Information Processing Systems}, 33:\penalty0 13199--13213, 2020.

\bibitem[Perrin et~al.(2022)Perrin, Lauri{\`e}re, P{\'e}rolat, {\'E}lie, Geist, and Pietquin]{Perrin2022General}
S.~Perrin, M.~Lauri{\`e}re, J.~P{\'e}rolat, R.~{\'E}lie, M.~Geist, and O.~Pietquin.
\newblock Generalization in mean field games by learning master policies.
\newblock In \emph{Proceedings of the AAAI Conference on Artificial Intelligence}, volume~36, pages 9413--9421, 2022.

\bibitem[Rashid et~al.(2020)Rashid, Samvelyan, De~Witt, Farquhar, Foerster, and Whiteson]{rashid2020monotonic}
T.~Rashid, M.~Samvelyan, C.~S. De~Witt, G.~Farquhar, J.~Foerster, and S.~Whiteson.
\newblock Monotonic value function factorisation for deep multi-agent reinforcement learning.
\newblock \emph{The Journal of Machine Learning Research}, 21\penalty0 (1):\penalty0 7234--7284, 2020.

\bibitem[Robinson(1951)]{robinson1951iterative}
J.~Robinson.
\newblock An iterative method of solving a game.
\newblock \emph{Annals of mathematics}, pages 296--301, 1951.

\bibitem[Schaul et~al.(2015)Schaul, Quan, Antonoglou, and Silver]{schaul2015prioritized}
T.~Schaul, J.~Quan, I.~Antonoglou, and D.~Silver.
\newblock Prioritized experience replay.
\newblock \emph{arXiv preprint arXiv:1511.05952}, 2015.

\bibitem[Shalev-Shwartz(2012)]{shalev2012online}
S.~Shalev-Shwartz.
\newblock Online learning and online convex optimization.
\newblock \emph{Foundations and Trends{\textregistered} in Machine Learning}, 4\penalty0 (2):\penalty0 107--194, 2012.

\bibitem[Subramanian and Mahajan(2019)]{subramanian2019reinforcement}
J.~Subramanian and A.~Mahajan.
\newblock Reinforcement learning in stationary mean-field games.
\newblock In \emph{Proceedings of the 18th International Conference on Autonomous Agents and MultiAgent Systems}, pages 251--259, 2019.

\bibitem[Vieillard et~al.(2020{\natexlab{a}})Vieillard, Kozuno, Scherrer, Pietquin, Munos, and Geist]{vieillard2020leverage}
N.~Vieillard, T.~Kozuno, B.~Scherrer, O.~Pietquin, R.~Munos, and M.~Geist.
\newblock Leverage the average: an analysis of kl regularization in reinforcement learning.
\newblock \emph{Advances in Neural Information Processing Systems}, 33:\penalty0 12163--12174, 2020{\natexlab{a}}.

\bibitem[Vieillard et~al.(2020{\natexlab{b}})Vieillard, Pietquin, and Geist]{vieillard2020munchausen}
N.~Vieillard, O.~Pietquin, and M.~Geist.
\newblock Munchausen reinforcement learning.
\newblock \emph{Advances in Neural Information Processing Systems}, 33:\penalty0 4235--4246, 2020{\natexlab{b}}.

\end{thebibliography}

\newpage
\appendix
\onecolumn
\section{Additional algorithm framework}
See Algo. \ref{algorithm2} for classic fictitious play (FP) and Algo. \ref{algorithm3} for classic online mirror descent (OMD).

\begin{algorithm}[htbp]
\caption{Classic Fictitious Play (FP)}\label{algorithm2}
\SetAlgoLined
\SetKwData{Left}{left}\SetKwData{This}{this}\SetKwData{Up}{up}
\SetKwFunction{Union}{Union}\SetKwFunction{FindCompress}{FindCompress}
\SetKwInOut{Input}{input}\SetKwInOut{Output}{output}
\Input{Number of iterations $K$,Initialize $\pi^0$} 
\begin{flushleft}
\For{$k=0, \ldots, K$}{ 

Forward Update: Compute $\mu^k = \mu^{\pi^{k-1}}$

Average Distribution Update: Compute $\bar{\mu}^k$ as the average of $(\mu^0,\ldots,\mu^{\pi^k})$ for every timestep $n$: 

\begin{equation*}\bar{\mu}_n^k(x) = \frac{1}{k}\sum_{i=1}^k \mu_n^i(x) = \frac{k-1}{k}\bar{\mu}^{k-1}_{n}(x) + \frac{1}{k}{\mu}_n^k(x)\end{equation*}

Best Response Computation: Compute a BR $\pi^k$ against $\bar{\mu}^k$, e.g. by computing $Q^{*,\bar{\mu}^k}$ and then taking $\pi_n^k(. \mid x)$ as a(ny) distribution over $\arg\max Q^{*,\bar{\mu}^k}(x,\cdot)$ for every $n,x$;
}

\Output{$\bar{\mu}^K = (\bar{\mu}_{n}^K)_{n=0,\ldots,NT}$ and policy $\bar{\pi}^K = (\bar{\pi}_{n}^K)_{n=0,\ldots,NT}$ generating this mean field sequence.} 
\end{flushleft}
\end{algorithm}

\begin{algorithm}[htbp]
\caption{Classic Online Mirror Descent (OMD)}\label{algorithm3}
\SetAlgoLined
\SetKwData{Left}{left}\SetKwData{This}{this}\SetKwData{Up}{up}
\SetKwFunction{Union}{Union}\SetKwFunction{FindCompress}{FindCompress}
\SetKwInOut{Input}{input}\SetKwInOut{Output}{output}
\begin{flushleft}
\textbf{Input:} {learning rate parameter $\tau$; number of iterations $K$; timestep $n$; \\
Initialize Q table $\left(\bar{q}_n^0\right)_{n=0, \ldots, N_T}$, e.g. with $\bar{q}_n^0(x, a)=0$ for all $n, x, a$} 
\end{flushleft}
\begin{flushleft}
\textbf{Output:} {policy $\pi_K$ and final regularized ${\bar{q}^K}$} 
Initialize:$(\bar{q}_n^0){n=0,\ldots,N_T}$, e.g. with $\bar{q}_n^0(x, a)=0$, for all $n$, $x$, $a$.
\end{flushleft}
\begin{flushleft}
Let the projected policy be: $\pi_n^0(a \mid x)=\operatorname{softmax}(\bar{q}_n^0(x,\cdot))(a)$ for all $n$, $x$, $a$.

\For{$k=1, \ldots, K$}{ 

Forward Update: $\mu^k=\mu^{\pi^{k-1}}$.

Backward Update: $Q^k=Q^{\pi^{k-1}, \mu^k}$.

Update the regularized  $Q$:\;

\qquad $\bar{q}_n^k(x, a)=\bar{q}_n^{k-1}(x, a)+\frac{1}{\tau} Q_n^k(x, a)$\;

\qquad $\pi_n^k(a \mid x)=\operatorname{softmax}(\bar{q}_n^k(x,\cdot))(a)$\;

}
\Return{$(\bar{q}^K, \pi^K)$}
\end{flushleft}
\end{algorithm}

\begin{algorithm} [htb]  
\caption{Master Deep Online Mirror Descent (detailed)}\label{algo: algorithm2}
\SetAlgoLined
\SetKwData{Left}{left}\SetKwData{This}{this}\SetKwData{Up}{up}
\SetKwFunction{Union}{Union}\SetKwFunction{FindCompress}{FindCompress}
\SetKwInOut{Input}{input}\SetKwInOut{Output}{output}

\begin{flushleft}
  \textbf{Input:} Learning iteration ${K}$; training episodes ${N_{episodes}}$ per iteration; Replay buffer $\mathcal{M_{RL}}$; horizon $N_T$, number of agents ${N}$; Initial distribution $\mathcal{D}$; Munchausen parameter $\tau$, Initial Q-network parameter $\theta$; Initial target Q-network parameter $\theta^{\prime}$
\end{flushleft}

\begin{flushleft}
\textbf{Output:}  {Policy $\boldsymbol{\pi}$}
\end{flushleft}

\begin{flushleft}
Set initial target network parameter $\theta^{\prime}$ = $\theta$ 

Set initial 
$\boldsymbol{\pi}^0(a \mid(n, x,\mu))=\operatorname{softmax}\left(\frac{1}{\tau} {Q}_{\theta}((n, x,\mu), \cdot)\right)$
\end{flushleft}
\begin{flushleft}
\For{iteration $k=1,2,\dots,K$} 
{
1. Update mean-field sequence:

\For{ distribution ${\boldsymbol{\mu}^k}$ in $(\boldsymbol{\mu}^{k,\mu_0})_{\mu_0 \in \mathcal{D}}$}
{
Update mean-field sequence $\boldsymbol{\mu}^{k}$ with $\boldsymbol{\pi}^{k-1}$ sampled by agents ${N}$ 
}
2. Reset the replay buffer $\mathcal{M_{RL}}$ 

3. Value function update:

\For{episode $t=1,2,\dots,N_{episodes}$}
 {
 \For{  ${\boldsymbol{\mu}^k}$ in $(\boldsymbol{\mu}^{k,\mu_0})_{\mu_0 \in \mathcal{D}}$}
 {
     \For{time step $n=1,2,\dots,N_T$}
        {
            Sample action $a_{n}$ from $\epsilon$-greedy policy based on $\tilde{Q}_{\theta}$ 
            
            Execute action $a_{n}$ and get the transition: 
            $\left\{\left(\left(n, x_{n}, \mu_{n}^k\right), a_n, r_n,\left(n+1, x_n^{\prime}, \mu_{n+1}^k\right)\right)\right\}$
            
            Store transition in replay buffer $\mathcal{M_{RL}}$ 
    
            Periodically update $\theta$ with one step gradient step using a minibatch $N_B$ from $\mathcal{M_{RL}}$: 
            $\theta \mapsto \frac{1}{N_B} \sum_{i=1}^{N_B}\left|\tilde{Q}_\theta\left(\left(n_i, x_{n_i},\mu_{n_i}^k\right), a_{n_i}\right)-T_{n_i}\right|^2$
            
            where $T$ is defined in \eqref{eq:target-Tn} 
            
            Periodically update target network parameter $\theta^{\prime} = \theta$
        }
 }
 
 }
    
4. Policy update: 
$\boldsymbol{\pi}^k(\cdot \mid(n, x,\mu))=\operatorname{softmax}\left(\frac{1}{\tau} \tilde{Q}_{\theta}((n, x,\mu), \cdot)\right)$
} 
Return $\boldsymbol{\pi}^K$
\end{flushleft}
\end{algorithm}

\section{More details on the Q-function updates}

\subsection{Equivalent formulation of the Q-function updates}
\label{proof:thm}
In the main text, we propose a Theorem~\ref{thm:equivalence} that is the key foundation of our algorithm as well as the corresponding proof. We denote by $\mathrm{KL}$ the Kullback-Leibler divergence: for $\pi_1,\pi_2 \in \Delta_{\mathcal{A}}$, $\mathrm{KL}\left(\pi_1 \| \pi_2\right)=\left\langle\pi_1, \ln \pi_1-\ln \pi_2\right\rangle$, which is also used in other regularized RL algorithms to make the training stage more stable, see e.g. \citep{vieillard2020munchausen,vieillard2020leverage,cui2021approximately,lauriere2022scalable}. By this theorem, we can give the $Q$ update and $\widetilde{Q}^k$ update equations respectively: 

 
\begin{equation}
\label{oldq}
    {Q}_n^k(x, \mu_n^{k},\cdot)=r_n(x, \mu_n^{k})+\gamma  \sum_{a'} 
    \pi_{n+1}^k (x, \mu_{n+1}^{k},a') Q^k_{n+1}(x, \mu_{n+1}^{k},a')
\end{equation}

\begin{equation}
\label{newq}
\begin{aligned}
     \tilde{Q}_n^k\left(x, \mu_n^k, a \right)&=r_n\left(x, \mu_n^k, a\right)  +\tau \ln \pi_n^{k-1}\left(x, \mu_n^k, a\right) \\
    & \qquad +\gamma \sum_{a'} 
    \pi_{n+1}^k\left(x, \mu_{n+1}^k, a'\right) \Bigg[\tilde { Q }_{ n + 1}^{ k }(x, \mu_{n+1}^k, a')\\
    &\qquad\qquad - \tau \ln \pi_{n+1}^{k-1}\left(x, \mu_{n+1}^k, a'\right) \Bigg]
\end{aligned}
\end{equation}

The main idea behind Theorem~\ref{thm:equivalence} is that we establish the connection between the two equations below. We first prove the two equalities~\eqref{softmax1} and \eqref{softmax2}, then prove the equivalence between the right hand side of both equations. See Appendix~
\ref{proof:softmax}. 
\begin{equation}
\label{softmax1}
     \operatorname{softmax}\left(\frac{1}{\tau}\sum_{i=0}^k Q^i\right)=\underset{\pi}{\operatorname{argmax}}\left\langle\pi, Q^k\right\rangle-\tau \mathrm{KL}\left(\pi \| \pi^{k-1}\right)
\end{equation}
\begin{equation}
\label{softmax2}
    \operatorname{softmax}\left(\frac{1}{\tau} \tilde{Q}^k\right) = \underset{\pi}{\operatorname{argmax}}\left\langle\pi, \tilde{Q}^k\right)-\tau\langle\pi , \ln \pi\rangle 
\end{equation}
Here we use the $\mathrm{softmax}$ instead of solving $\mathrm{argmax}$ in~\eqref{softmax2}, because solving  $\mathrm{argmax}$ is not always guaranteed when using Deep RL, and the $\mathrm{softmax}$ implicitly contains the greedy step which can be exactly computed, which also benefits the convergence \citep{vieillard2020leverage}.

Therefore, for the cost function of Q update in Algo. \ref{algo: algorithm1}: 
\begin{equation}
    \mathbb{E}\left|\tilde{Q}^k_\theta\left(\left(n, x_n,\mu^k_n\right), a_n\right)-T\right|^2
\end{equation}
where $T$ is defined in \eqref{eq:target-Tn}.

Differing from the Q update function proposed in \citep{lauriere2022scalable}, where the target policy is set as the policy learned from the previous iteration, our algorithm employs the target policy and target Q-function being learned in the current iteration. This modification helps to improve the convergence, which can be understood as follows: If the target policy remains fixed by a separate policy, the distribution of the policy under evaluation would differ from the one being learned. Consequently, this distribution shift would potentially introduce instability in the learning process. 

\subsection{Proof of Theorem~\ref{thm:equivalence}}
\label{proof:softmax}

\begin{proof}

In order to prove the result,  we will first expand the expressions for $\operatorname{softmax}\left(\frac{1}{\tau} \sum_i^k Q^i\right)$ and $\operatorname{softmax}\left(\frac{1}{\tau} \widetilde{Q}^k\right)$ to obtain~\eqref{softmax1} and~\eqref{softmax2}. Then, we will show that the right-hand sides of~\eqref{softmax1} and~\eqref{softmax2} are equivalent.


\noindent{\bf Step 1.} We prove the expansion of~\eqref{softmax1}, i.e. 
\begin{equation}
\begin{aligned}
    &\operatorname{softmax}\left(\frac{1}{\tau} \sum_{i=0}^k Q^i(x,\mu^k_n,n)\right)
    \\
    &=\underset{\pi}{\operatorname{argmax}}\left\langle\pi, Q^k(x,\mu^k_n,n)\right\rangle- \\
    &\qquad\tau \mathrm{KL}\left(\pi(\cdot \mid x,\mu^k_n,n) \| \pi^{k-1}(\cdot \mid x,\mu^k_n,n)\right)
\end{aligned}
\label{eq16}
\end{equation}

Here $Q^k$ is the standard Q-function at iteration $k$, as defined in~\eqref{oldq}. $\pi^{k-1}$ is the policy learned in iteration $k-1$.  $\mu^{k}$ is the mean field induced by the policy $\pi^{k-1}$. 

First, we define a new function, denoted as $F$, which corresponds to the right-hand side of \eqref{eq16}. For brevity, we exclude the explicit inputs of $\pi$ and $Q$, since the optimization process solely focuses on optimizing the parameter $\pi$. Hence, we express this simplification as $Q = Q(x,\mu_n,n)$. Since the policy is the probability distribution, an additional constraint is needed to guarantee that the sum of $\pi$ is 1.  
\begin{equation}
\begin{gathered}
\underset{\pi}{\operatorname{argmax}} \, F(\pi)=\underset{\pi}{\operatorname{argmax}}\left\langle\pi, Q^k\right\rangle-\tau \mathrm{KL}\left(\pi \| \pi^{k-1}\right) \\
\text { s.t. } \quad \boldsymbol{1}^{\top} \pi=1,
\end{gathered}
\label{prob:optimization1}
\end{equation}
where $\boldsymbol{1}$ denotes a vector full of ones, of dimension the number of actions.

We then introduce the Lagrange multiplier $\lambda$ and the Lagrangian $L$, defined as:
\begin{equation}
    L(\pi,\lambda)=\left\langle\pi, Q^k\right\rangle-\tau \mathrm{KL}\left(\pi \| \pi^{k-1}\right)+\lambda\left(\boldsymbol{1}^{\top} \pi-1\right)
\label{eq:lagrange}
\end{equation}

Now, by finding the equilibrium, we need to find a saddle point of~\eqref{eq:lagrange}. So let us compute the partial derivatives of $L$. Proceeding formally, we obtain:
\begin{equation}
    \begin{aligned}
\frac{\partial L(\pi, \lambda)}{\partial \pi} & =Q^k-\tau \frac{\partial\langle\pi \ln \pi\rangle}{\partial \pi}+\tau \frac{\left\langle\pi, \ln \pi^{k-1}\right\rangle}{\partial \pi}+\lambda(\mathbf{1}) \\
& =Q^k-\tau\left(\ln \pi+1-\ln \pi^{k-1}\right)+\lambda \mathbf{1} \\
\frac{\partial L(\pi, \lambda)}{\partial \lambda} & =\mathbf{1}^{\top} \pi-1
\end{aligned}
\label{eq:second_pd}
\end{equation}

Note that, for every $\lambda$, the function $\pi \mapsto L(\pi,\lambda)$ is concave, as the sum of a linear function and the negative of the KL divergence (and since the KL divergence is convex). 
Taking $\frac{\partial L(\pi,\lambda)}{\partial \pi}=0$, we find that the optimum satisfies:
\begin{equation}
\begin{aligned}
\pi & =e^{\frac{1}{\tau} (Q^k +\lambda\boldsymbol{1}) -1} \cdot \pi^{k-1} \\
& =e^{\frac{1}{\tau} (Q^k +\lambda\boldsymbol{1})-1} \cdot e^{\frac{1}{\tau} \cdot (Q^{k-1} +\lambda\boldsymbol{1})-1} \ldots \\
& =e^{\frac{1}{\tau}\left(Q^k+Q^{k-1}+\cdots+Q^0+(k+1)\lambda\boldsymbol{1}\right)-(k+1)} \\
& =e^{\frac{1}{\tau} \sum_{i=0}^k Q^i} \cdot e^{-(k+1)} \cdot e^{\frac{(k+1)}{\tau} \lambda\boldsymbol{1}} \\
& =e^{\frac{1}{\tau} \sum_{i=0}^k Q^i}\cdot {C_1} \cdot {C_2{(\lambda})}
\end{aligned}
\label{eq:first_pd2}
\end{equation}
where $C_1$ is a constant which equals to $e^{-(k+1)}$, $C_2$ is the function of the Lagrange multiplier $\lambda$. In order to satisfy the constraint  $\frac{\partial L(\pi,\lambda)}{\partial \lambda}=0$, i.e. $\operatorname{sum}(\pi)=\sum \pi=1$, note that $\lambda$ is a scalar, and $\frac{1}{\tau} \sum_{i=0}^k Q^i$ is a vector with dim of $|A|$, if \eqref{eq:lambda} holds, then the optimization problem \eqref{prob:optimization1} is solved. 

\begin{equation}
    C_1 \cdot C_2(\lambda) = \frac{1}{\sum_{\frac{1}{\tau} \sum_{i=0}^k Q^i} e^{\frac{1}{\tau} \sum_{i=0}^k Q^i}}
\label{eq:lambda}
\end{equation}

 
Thus, $\pi$ is a softmax function as follows, 
\begin{equation}
\begin{aligned}
\pi= \operatorname{softmax}\left(\frac{1}{\tau} \sum_{i=0}^k Q^i\right)
\end{aligned}
\end{equation}

\noindent{\bf Step 2. } Define a new function $G(\pi)$ 

\begin{equation}
    \underset{\pi}{\operatorname{argmax}} G(\pi) = \underset{\pi}{\operatorname{argmax}}\left\langle\pi, \tilde{Q}^k\right)-\tau\langle\pi , \ln \pi\rangle 
\end{equation}
following the same way as solving \eqref{prob:optimization1}, we can prove the second equality as needed. i.e the expansion of \eqref{softmax2}:  \
\begin{equation}
    \underset{\pi}{\operatorname{argmax}}\left\langle\pi, \tilde{Q}^k\right\rangle-\tau\langle\pi , \ln \pi\rangle =  \operatorname{softmax}\left(\frac{1}{\tau} \tilde{Q}^k\right) 
\end{equation}

\noindent{\bf Step 3. }
We now prove the equivalence between the two right hand sides. 
Let $\tilde{Q}^k=Q^k+\tau \ln \pi_{k-1}$.
Then 
\begin{equation}
    \begin{aligned}
&\underset{\pi}{\operatorname{argmax}}\left\langle\pi \cdot Q^k\right\rangle-\tau \mathrm{KL}\left\langle\pi \| \pi^{k-1}\right\rangle 
\\
& =\underset{\pi}{\operatorname{argmax}}\left\langle\pi, \tilde{Q}^k-\tau \ln \pi_{k-1}\right\rangle-\tau \mathrm{KL}\left(\pi \| \pi^{k-1}\right) \\
& =\underset{\pi}{\operatorname{argmax}}\left\langle\pi, \tilde{Q}^k\right\rangle-\tau\langle\pi, \ln \pi\rangle
\end{aligned}
\end{equation}
Therefore, Theorem~\ref{thm:equivalence} is proved.

\end{proof}

\section{More details about algorithms}

\subsection{Ad-hoc teaming}
\label{sec:adhoc-teaming}
\begin{figure}[htbp]
  \centering
  \subfloat[Ad-hoc teaming: Small team joining]{
    \includegraphics[width=0.45\columnwidth]{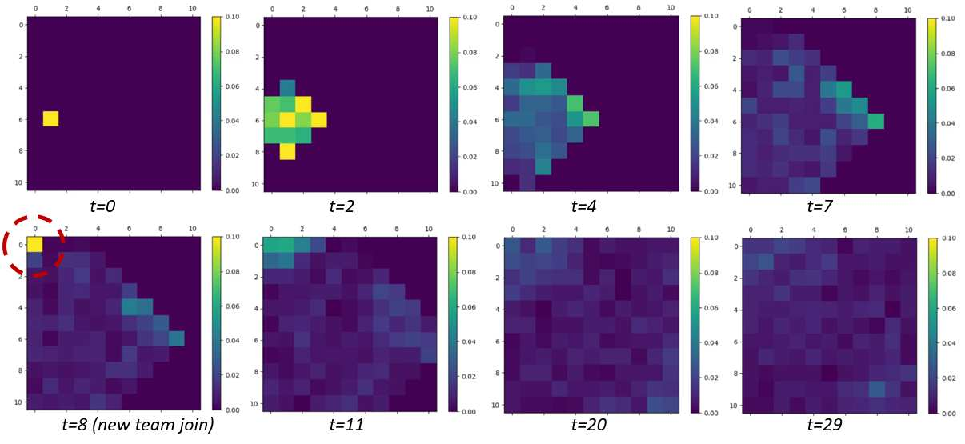}
  }
  \subfloat[Ad-hoc teaming: Large team joining]{
    \includegraphics[width=0.45\columnwidth]{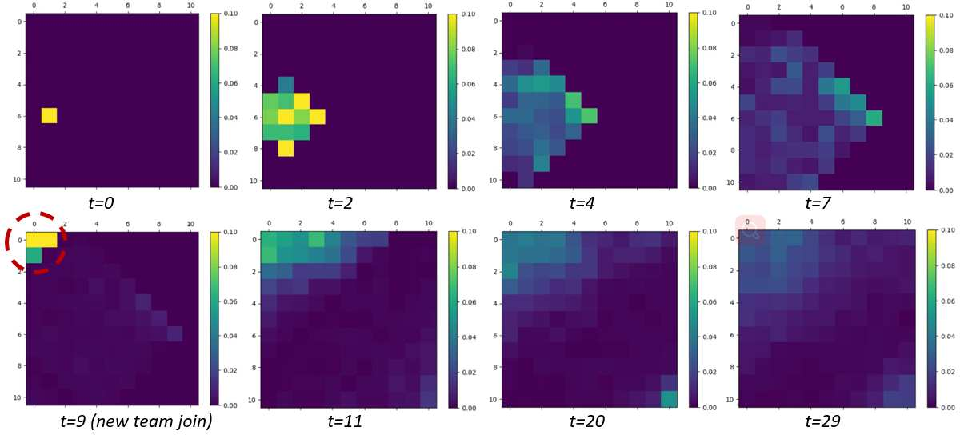}
  }
  \caption{Ad-hoc teaming test simulates new team joining into the current team (500 agents) during the evolutive process of the population. The simulation contains two different teams, one is a small team joining (200 agents), another is a large group team joining (2500 agents)}
  \label{fig:ad-hoc}
\end{figure}

In this paper, we introduce a novel testing case for the master policy, referred to as Ad-hoc teaming, inspired by ad-hoc networks in telecommunication. Ad-hoc teaming simulates the scenario where additional agent groups join the existing team during execution, resembling spontaneous and temporary formations without centralized control or predefined network topology. By incorporating the concepts of distribution and time awareness, the policy should enable multiple teams to join at any timestep, ultimately achieving the Nash equilibrium. Fig. \ref{fig:ad-hoc} illustrates two instances where different agent groups join the current team during execution. One group consists of a small number of agents, causing minimal impact on the overall population distribution, while the other group comprises a larger number of agents, significantly altering the distribution. As shown in Fig.~\ref{fig:ad-hoc}, the population still leads to uniform distribution after the small team joins. However, when a large group joins the current team, the final distribution at the terminal timestep deviates from the expected uniform distribution. The reasons can be attributed to two aspects. Firstly, the time left on the horizon is not sufficient to allow ad-hoc agents to spread out. Secondly, the generalization learning limits. During the training process, the population initially starts from a single area and subsequently spreads across the map. Consequently, the policy networks' distribution awareness is implicitly limited to the spread-out tendency. However, the random emergence of ad-hoc teams disrupts this spread-out distribution of the population. To address this issue, it is necessary to incorporate additional scenarios like ad-hoc teaming during training. This paper presents an initial exploration of such a testing scenario in real life, leaving a more comprehensive investigation for future research.

\begin{figure}[h!]      
  \centering
  \includegraphics[width=0.6\textwidth]{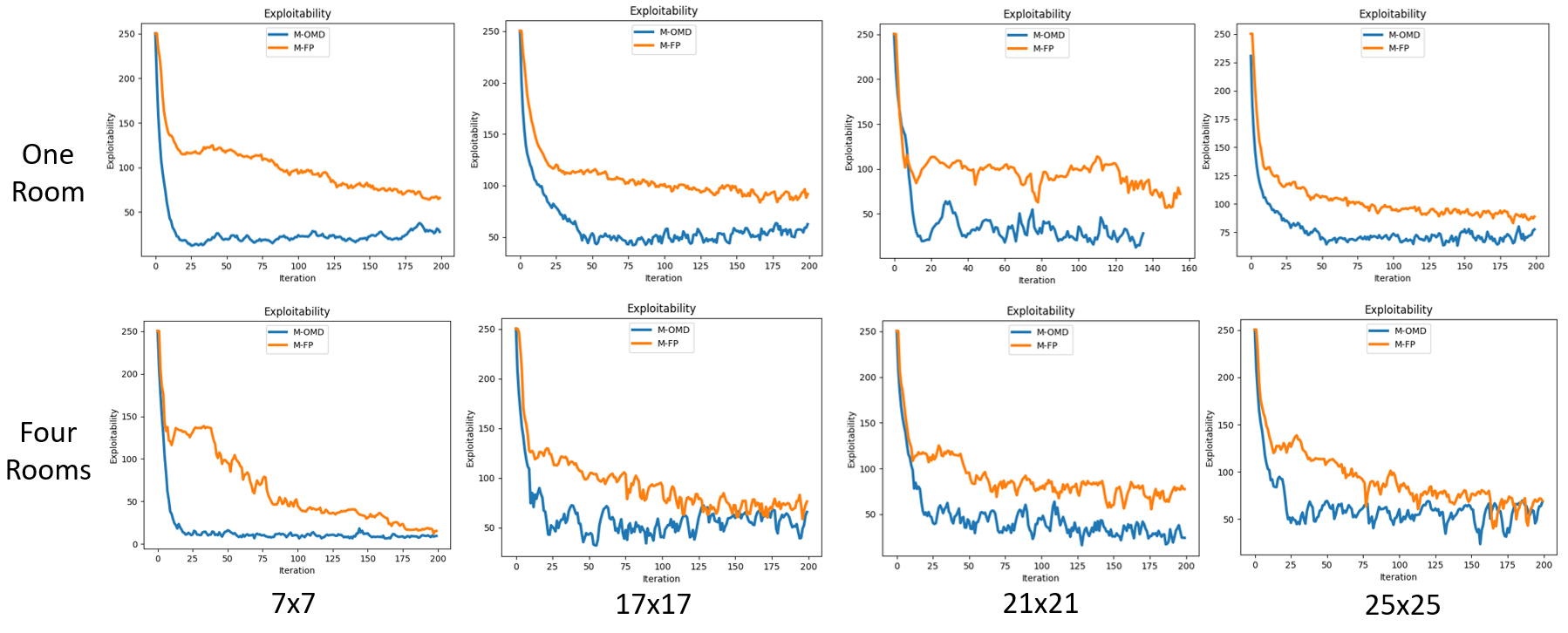}
  \caption{Exploitability versus map size. We compared two master policies: M-OMD (ours) and M-FP(SOTA) in five different map dimensions. Due to the time horizon of games, a map size larger than 25x25 is not meaningful as agents cannot explore even half the map before termination.}
  \label{fig:map_size}
\end{figure}

\begin{figure}[htbp]
  \centering
  \subfloat[Memory vs Map size]{
    \includegraphics[width=0.2\columnwidth]{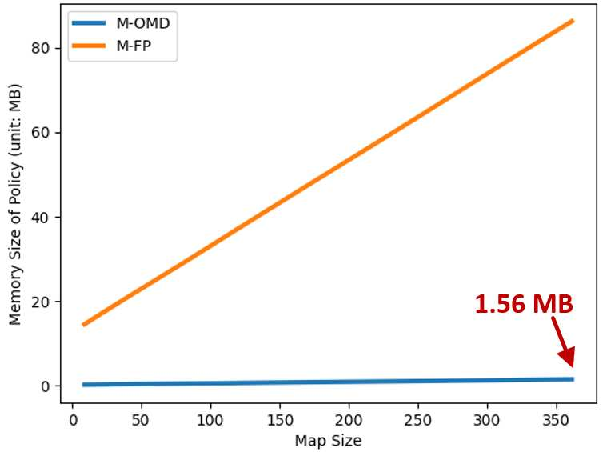}
  }
  \subfloat[Memory vs Iteration training]{
    \includegraphics[width=0.2\columnwidth]{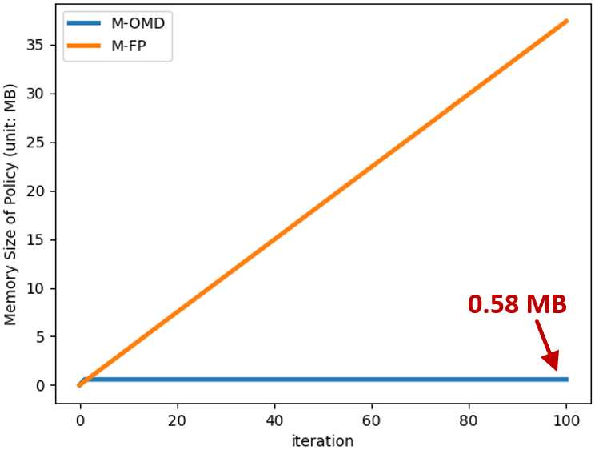}
  }
  \caption{Model size comparison for M-OMD (ours) and M-FP.}
  \label{memory_cost}
\end{figure}

\subsection{Training set and testing set}
\begin{figure}[htbp]
  \centering
  \subfloat[Training set]{
    \includegraphics[width=0.45\textwidth]{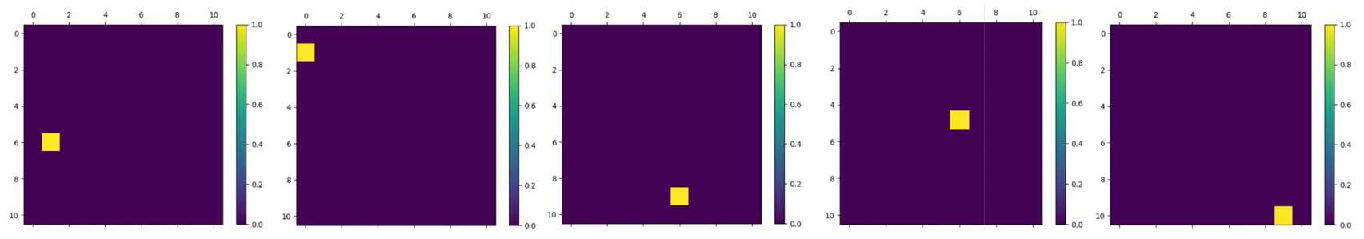}
  }
  \subfloat[Testing set]{
    \includegraphics[width=0.45\textwidth]{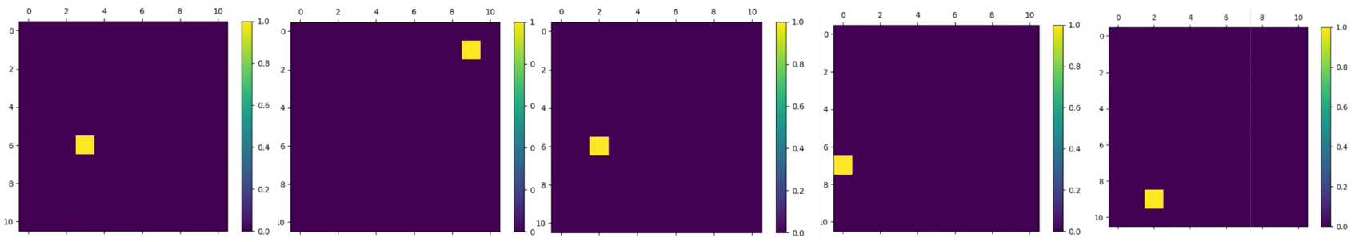}
  }
  \caption{Training and testing sets for Beach bar task \& Exploration in One room task}
  \label{fig:trainset_uniform}
\end{figure}

\begin{figure}[htbp]
  \centering
  \subfloat[Training set]{
    \includegraphics[width=0.45\textwidth]{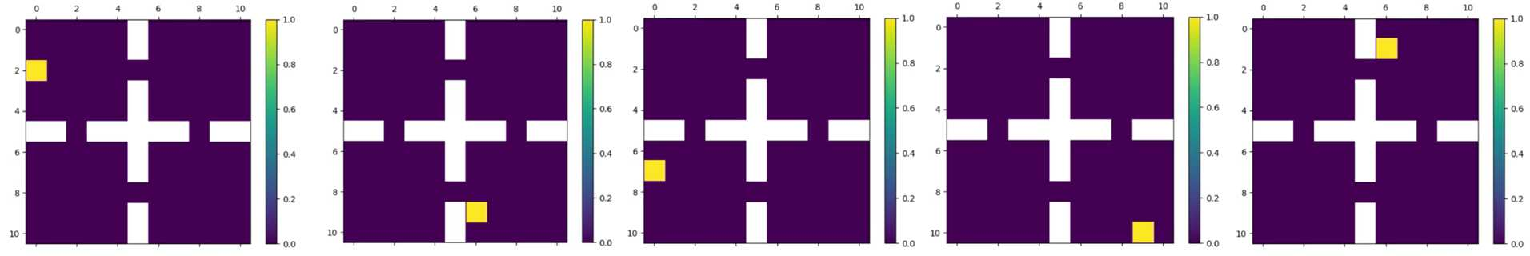}
  }
  \subfloat[Testing set]{
    \includegraphics[width=0.45\textwidth]{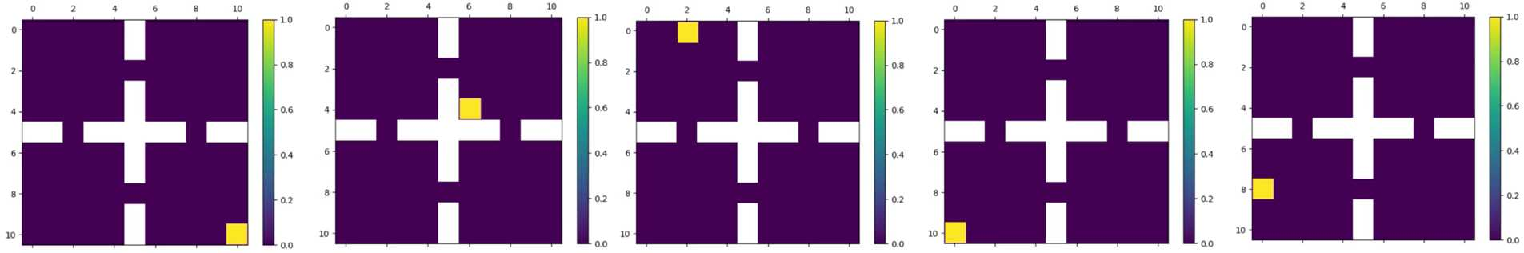}
  }
  \caption{Training and testing sets for Exploration in Four Rooms task}
  \label{fig:trainset_exploration2}
\end{figure}

To validate the effectiveness of the learned master policy, we adopt the approach described in \citep{Perrin2022General} to construct separate training and testing sets. This section presents the five training sets used to learn the policy and five testing sets utilized to evaluate its performance. The distributions for the Beach bar task and Exploration in one room task are depicted in Fig. \ref{fig:trainset_uniform}, while Fig. \ref{fig:trainset_exploration2} illustrates the distributions for the Exploration in four rooms task. We provide a summary of the exploitability of the testing sets in Table ~\ref{table:exploitability_testing}. It shows that all algorithms exhibit higher exploitability than the training set, as evidenced by the exploitability curves in the training figures at the final iteration. We attribute this to overfitting and insufficient training data, which are classic challenges in the field of machine learning. The M-OMD results demonstrate a significant reduction in exploitability during training, although it does not maintain the same level during testing. Insufficient data implies a limited representation of diverse initial distributions, causing the neural network to struggle with changes in population distribution, which is also revealed in the Ad-hoc teaming tests to some extent. However, it is important to note that overfitting and insufficient amounts of training data are common issues in ML, which do not undermine the feasibility of our algorithm. Addressing these challenges and improving the effectiveness of training are the topics for future research. The reason why vanilla policies perform better than master policies during testing has been discussed in the LQ section, see Section~\ref{section:lq}.

\subsection{Hyperparameters for experiments}
See table \ref{table:hyperparam} for the training parameters for all five algorithms.
\begin{table}[htbp]
\caption{Hyperparameters used for training Exploration task with map size:11x11}
\label{table:hyperparam}
\vskip 0.15in
\begin{center}
\begin{small}
\begin{sc}
\begin{tabular}{lccccc}
\toprule
Algorithm & M-OMD & M-FP & V-FP & V-OMD1 & V-OMD2 \\
\midrule
Neural Network Arch & mlp & mlp & mlp & mlp & mlp \\
Neurons in Each Layer & 64*64 & 64*64 & 64*64 & 64*64 & 64*64 \\
Environment Horizon & 30 & 30 & 30 & 30 & 30 \\
Number of Agents & 500 & 500 & 500 & 500 & 500 \\
Max Steps per iteration & 30000 & 30000 & 30000 & 30000 & 30000 \\
OMD $\tau$ & 50 & N/A & N/A & 5.0 & 50\\
OMD $\alpha$ & N/A & N/A & N/A & 1.0 & 1.0\\
Freq to update target & 4 & 4 & 4 & 4 & 4 \\
Exploration Fraction & 0.1 & 0.1 & 0.1 & 0.1 & 0.1 \\
$\gamma$ & 0.99 & 0.99 & 0.99 & 0.99 & 0.99 \\
Batch Size & 32 & 32 & 32 & 32 & 32 \\
Gradient Steps & 1 & 1 & 1 & 1 & 1 \\
\bottomrule
\end{tabular}
\end{sc}
\end{small}
\end{center}
\vskip -0.1in
\end{table}

\subsection{Hyperparameter sweeping}
\begin{figure}[htbp]
  \centering
  \subfloat[M-OMD]{
    \includegraphics[width=0.45\columnwidth]{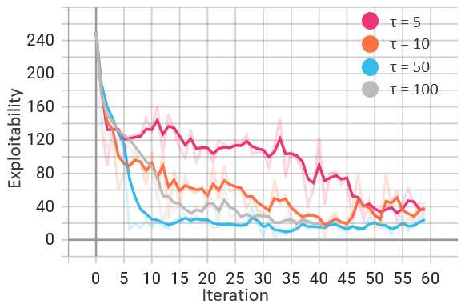}
  }
  \hfill 
  \subfloat[V-OMD1]{
    \includegraphics[width=0.45\columnwidth]{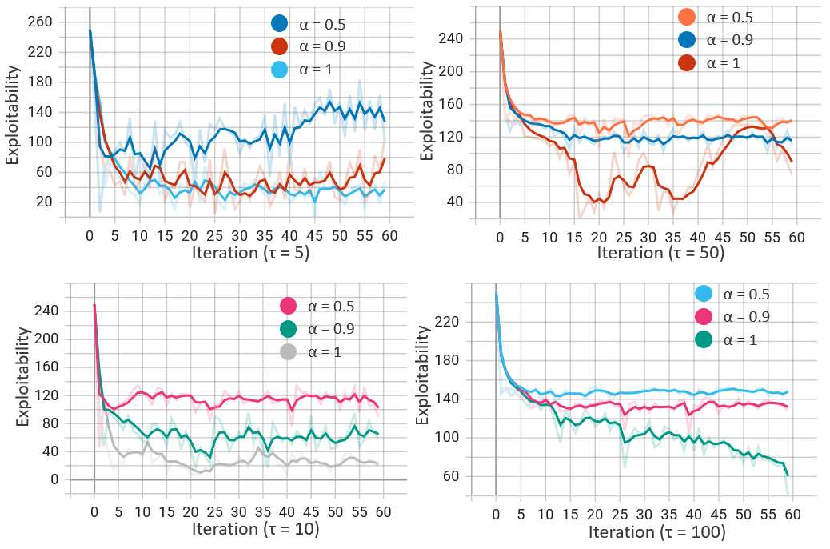}
  }
  \caption{Hyperparameter sweeping}
  \label{fig: sweep}
\end{figure}

We provide the sweeping curves of hyperparameter $\tau$, both ours and V-OMD1, and $\alpha$, only in V-OMD1 \citep{lauriere2022scalable} in this section. See Fig. \ref{fig: sweep}.

\subsection{Computational time}
\label{sec:comput-time}
\begin{figure}[htb]
\centering
\subfloat[Total time per iteration]{
\begin{minipage}[t]{0.2\linewidth}
\centering
\includegraphics[width=1.44in]{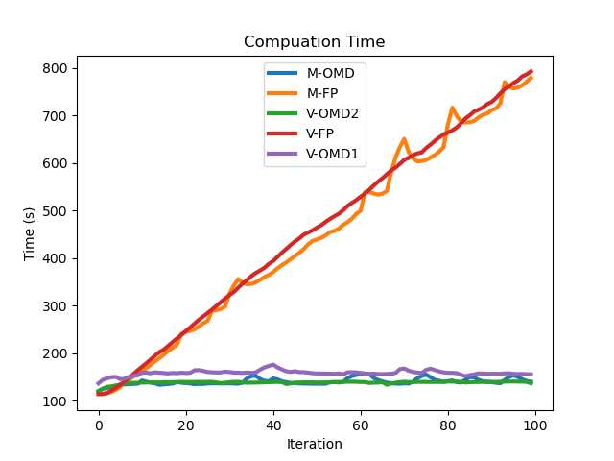}
\end{minipage}%
}%
\subfloat[Learn policy]{
\begin{minipage}[t]{0.2\linewidth}
\centering
\includegraphics[width=1.45in]{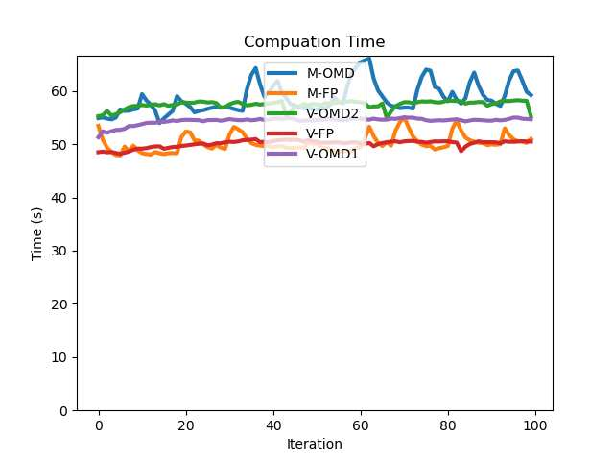}
\end{minipage}%
}%
\subfloat[Update distribution]{
\begin{minipage}[t]{0.2\linewidth}
\centering
\includegraphics[width=1.45in]{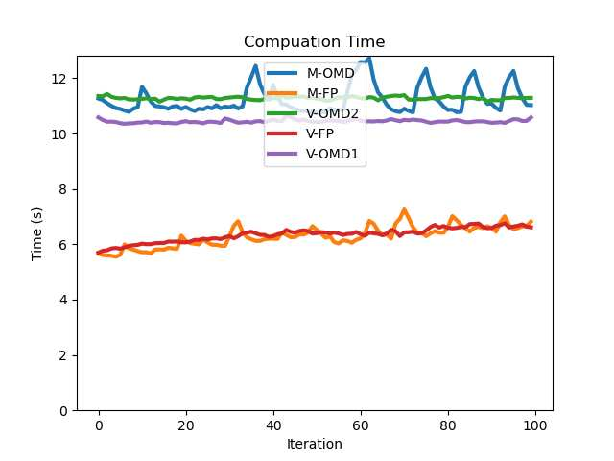}
\end{minipage}%
}%
\subfloat[Compute exploitability]{
\begin{minipage}[t]{0.2\linewidth}
\centering
\includegraphics[width=1.45in]{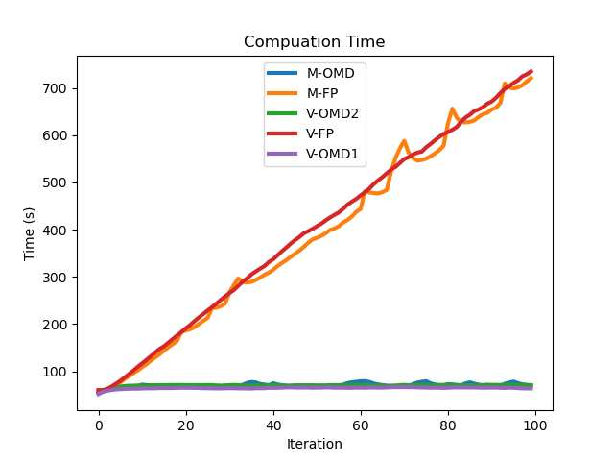}
\end{minipage}
}%
\caption{Computational time comparison (Exploration task). Figures correspond to three steps in each iteration, (b)learning the best response in FP or evaluating Q in OMD; (c) updating population distribution based on new learned policy;(d) calculating Exploitability}
\label{fig:time_computation}
\end{figure}

To calculate the exploitability during training, FP needs to use all history policies to execute while our algorithm only uses a single policy network, which results in a linear increase in computation cost for FP during training but not for our algorithm. Therefore, even though the policy learning time per iteration of our algorithm would be slightly longer than FP-based algorithms, it still saves much more computational time considering the convergence speed and computation of exploitability. See Fig.\ref{fig:time_computation} for details.

\subsection{Training with 30 initial distributions}
\begin{figure}[htbp]
  \centering
{
    \includegraphics[width=0.5\columnwidth]{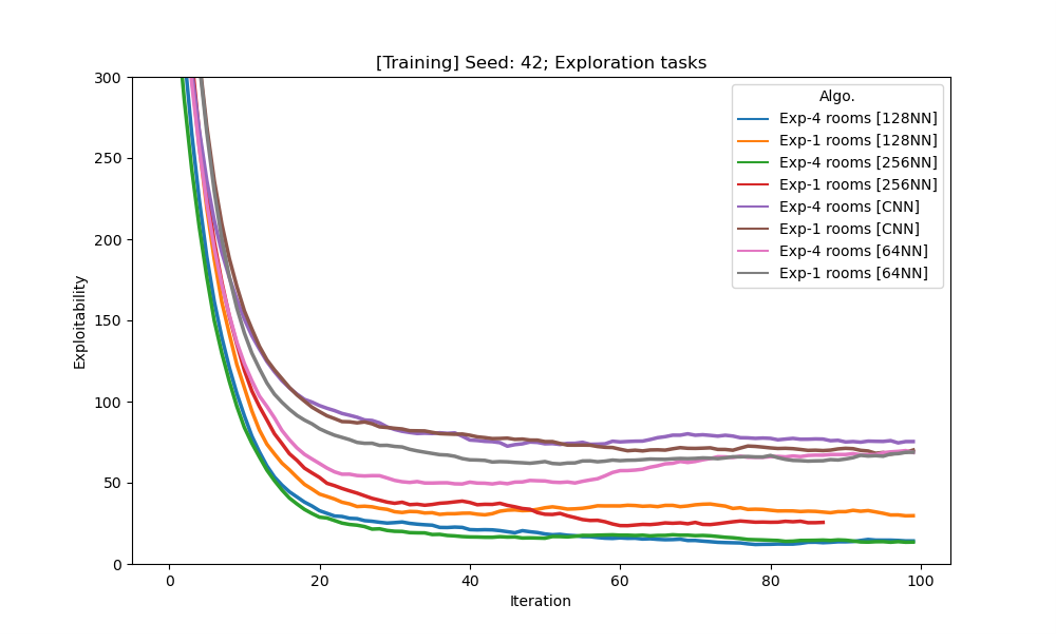}
  }
  \caption{Training on two exploration tasks with 30 distributions. Four different architectures are tested for Master OMD algorithm}
  \label{fig:exploitability_nn_archtecture}
\end{figure}

\begin{figure}[htb]
\centering
\subfloat[Exploration in 1 room (30 $\mu_0$) ]{
\includegraphics[width=0.45\linewidth]{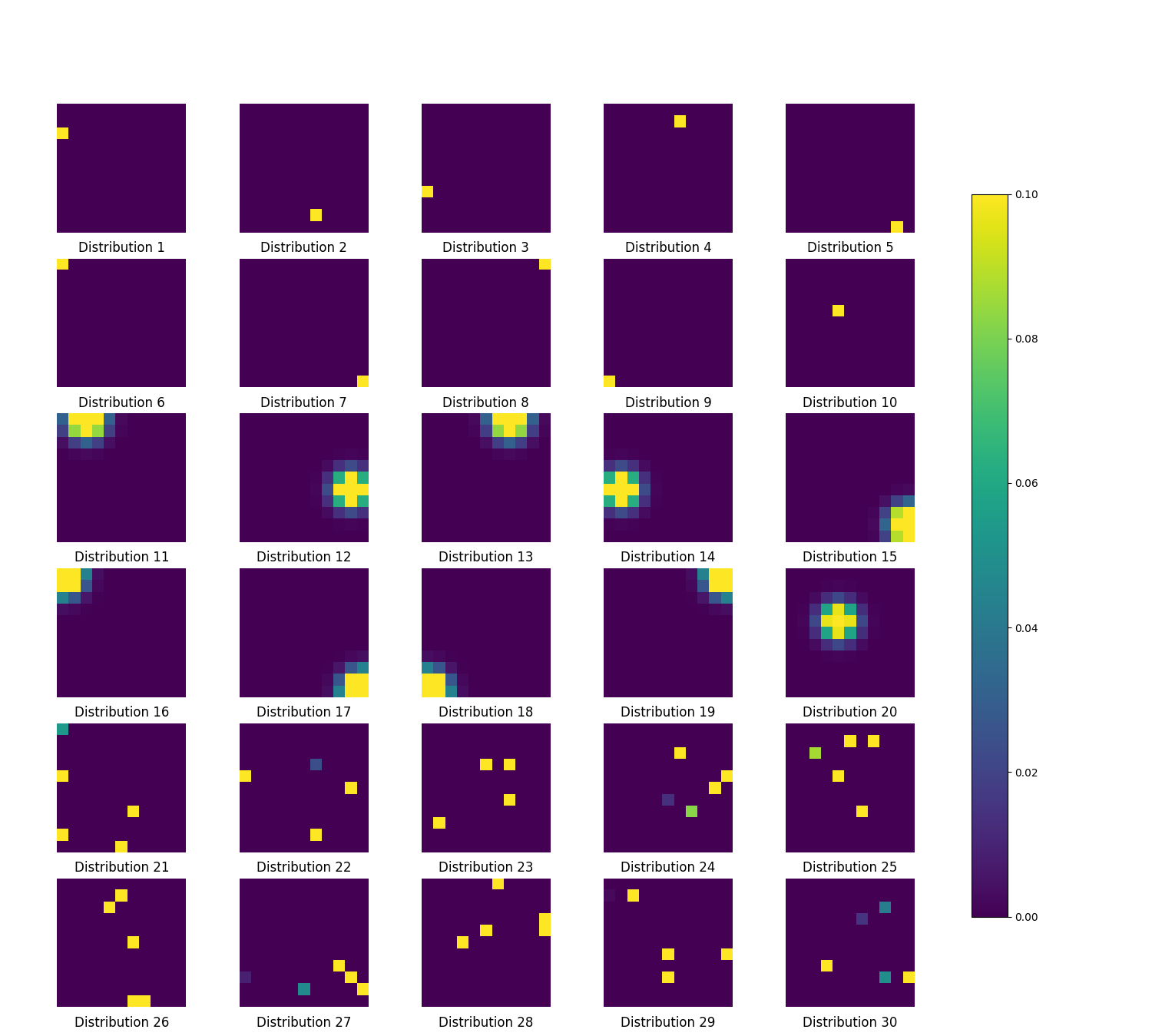}
}%
\subfloat[Exploration in 4 rooms (30 $\mu_0$) ]{
\includegraphics[width=0.45\linewidth]{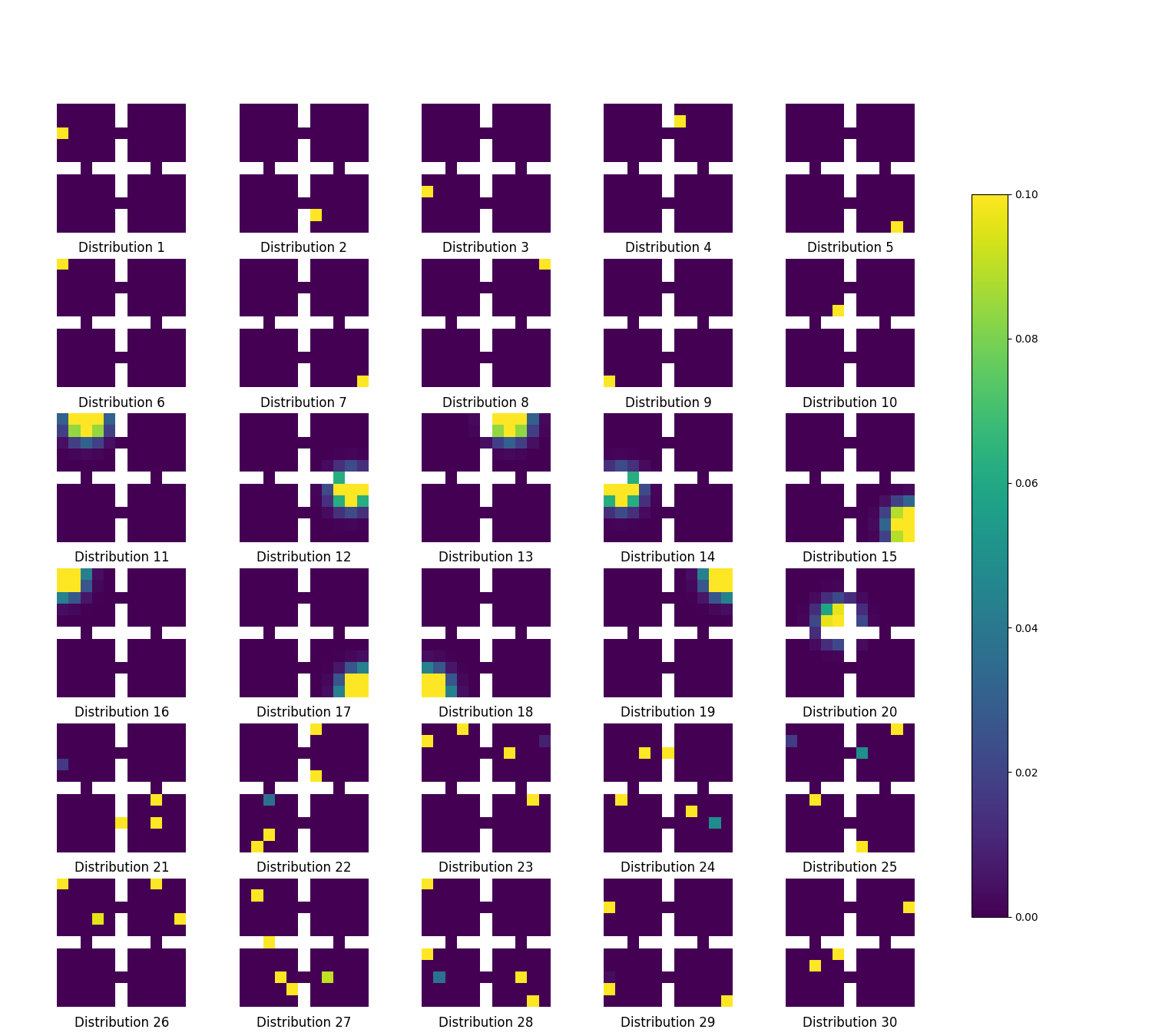}
}%

\caption{Training sets with 30 distributions for two exploration tasks (a) Exploration in One room (b) Exploration in Four room
} 
\label{fig:training_set_dis30}
\end{figure}

In the main text, we demonstrate that our algorithm proficiently handles five distributions, as illustrated in Fig \ref{fig:trainset_uniform}. To extend our exploration of its adaptability across a broader spectrum of distributions, we examine an ensemble of 30 distributions depicted in Fig \ref{fig:training_set_dis30}. This set comprises 10 distributions originating from fixed points, 10 following Gaussian distributions, and 10 distributed across random points. To assess the impact of policy architecture on performance, we evaluated four distinct architectures: three MLP-based architectures with configurations of $64\times64$, $128\times128$, and $256\times256$ layers, respectively, alongside a CNN-based architecture featuring two convolutional layers ($32\times64$, with kernel sizes of 5 and 3) followed by a fully connected layer. As evidenced in Fig \ref{fig:exploitability_nn_archtecture}, the $64\times64$ MLP architecture struggles to converge when handling 30 distributions. Consequently, we adopt the $256\times256$ architecture as our benchmark for further investigation. 

While the utilization of DQN for learning the optimal response is prevalent in Deep Mean Field Games (MFG), we aim to distinguish clearly between the DQN-derived best response and the authentic best response. To achieve this, we employ dynamic programming to solve Eq. \ref{eq:br_definition}, enabling us to precisely compute the true exploitability. We conducted experiments using both the Master Fictitious Play (M-FP) and Master Online Mirror Descent (M-OMD) algorithms, maintaining identical network architectures across two exploration tasks, each tested with five seeds: 42, 3407, 303, 109, and 312. As shown in Fig \ref{fig:exploitability_dis30}, our approach outperforms the population-based FP algorithm in terms of true exploitability, but also demonstrates advantages in computational efficiency, execution time, and model compactness, as previously highlighted.

\begin{figure}[htb]
\centering
\subfloat[Training stage: Exploration in One room (30 $\mu_0$) ]{
\includegraphics[width=0.45\linewidth]{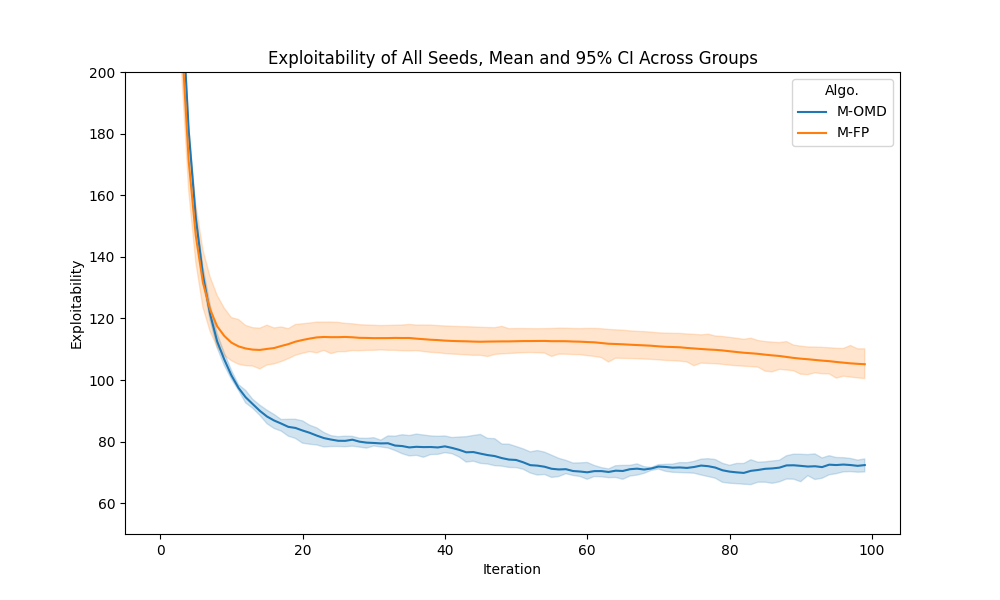}
}%
\subfloat[Testing stage: Exploration in Four rooms (30 $\mu_0$)]{
\includegraphics[width=0.45\linewidth]{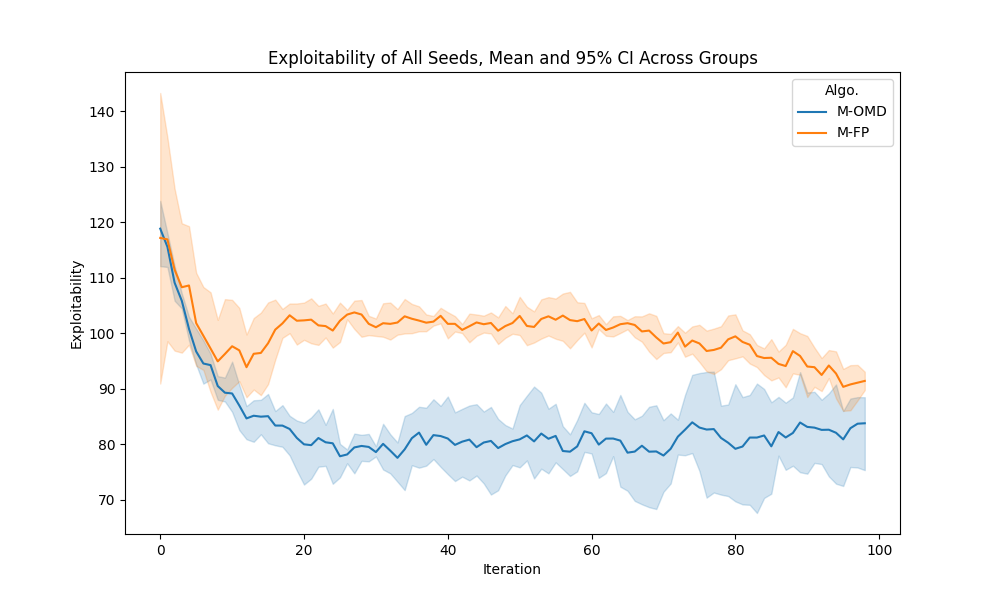}
}%

\caption{Exploration in One Room task: Exploitability for 30 initial distributions with the best response based on dynamic programming with MLP architecture $[256\times256]$. (a) is the comparison between Master FP and Master OMD (ours) in the training stage and (b) is the comparison between Master FP and Master OMD (ours) in the testing stage
} 
\label{fig:exploitability_dis30}
\end{figure}

\begin{figure}[htb]
\centering
\subfloat[Training stage: Exploration in Four rooms (30 $\mu_0$) ]{
\includegraphics[width=0.45\linewidth]{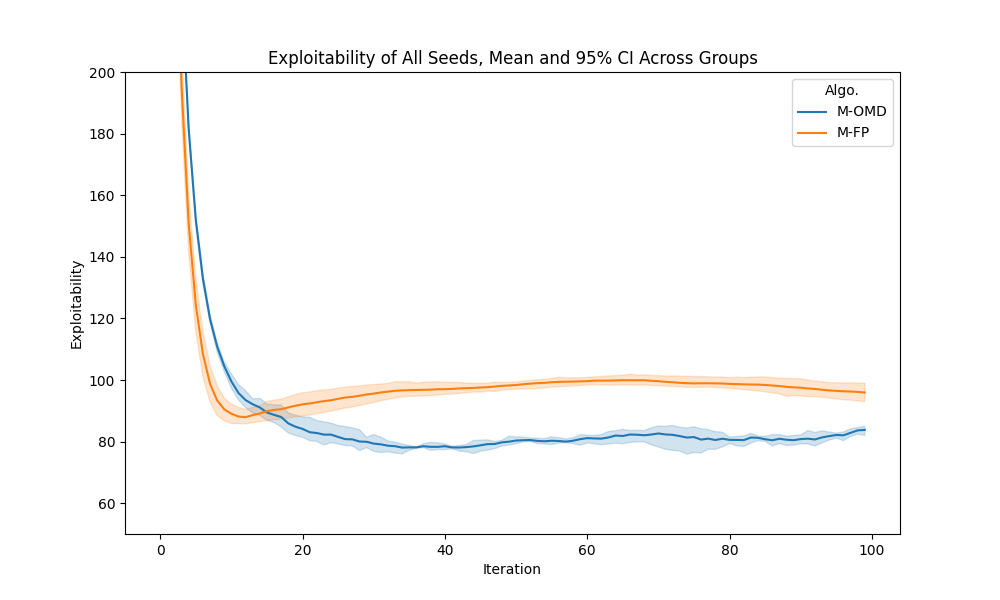}
}%
\subfloat[Testing stage: Exploration in Four rooms (30 $\mu_0$) ]{
\includegraphics[width=0.45\linewidth]{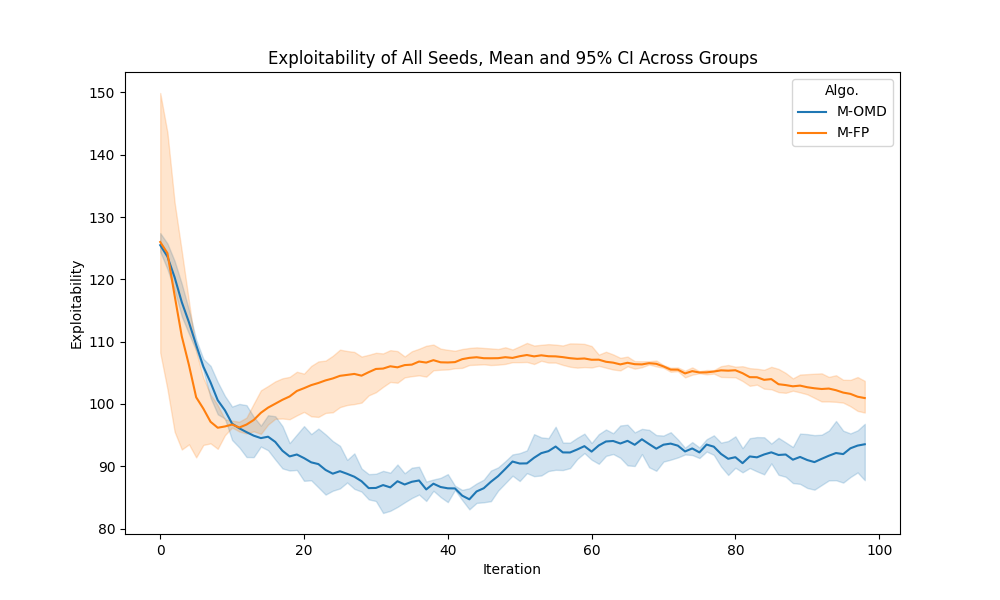}
}%

\caption{Exploration in Four Rooms task: Exploitability for 30 initial distributions with the best response based on dynamic programming with MLP architecture $[256\times256]$. (a) is the comparison between Master FP and Master OMD (ours) in the training stage and (b) is the comparison between Master FP and Master OMD (ours) in the testing stage
} 
\label{fig:exploitability_dis30}
\end{figure}

\begin{figure}[htb]
\centering
\subfloat[Exploration in 1 room (30 $\mu_0$) ]{
\includegraphics[width=0.45\linewidth]{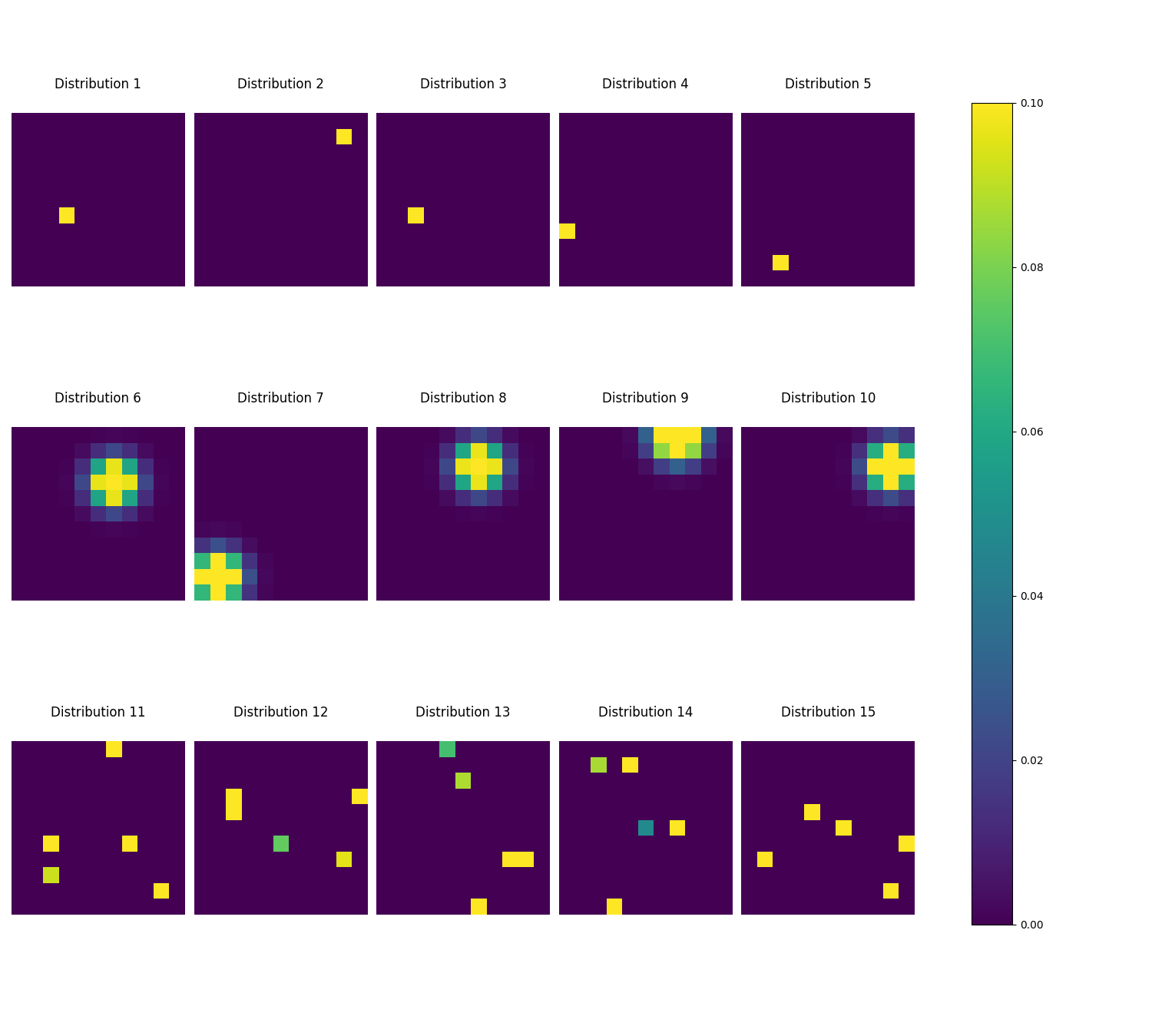}
}%
\subfloat[Exploration in 4 rooms (30 $\mu_0$) ]{
\includegraphics[width=0.45\linewidth]{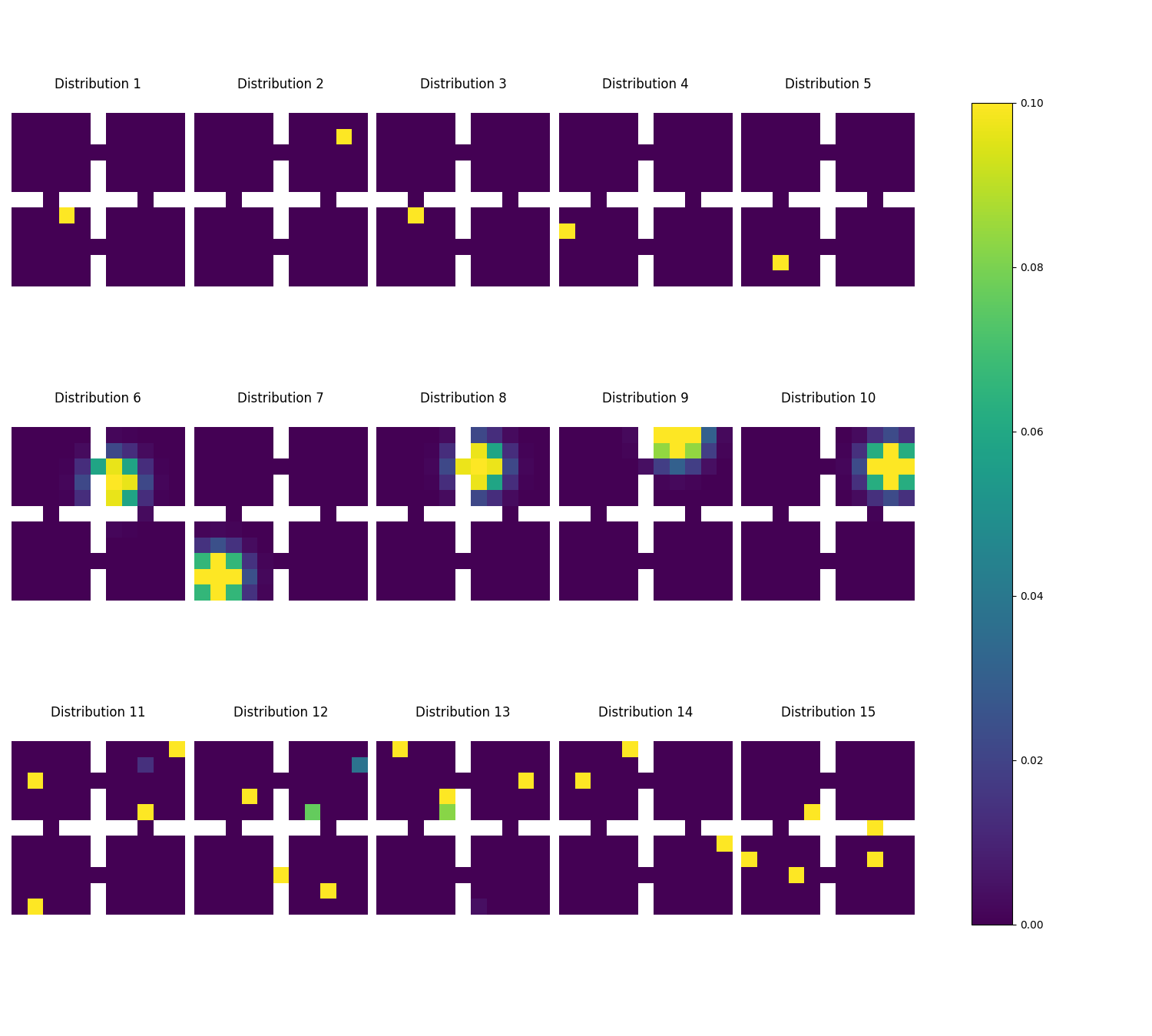}
}%

\caption{Testing sets with 30 distributions for two exploration tasks (a) Exploration in One room (b) Exploration in Four room
} 
\label{fig:tesing_set_dis30}
\end{figure}

\end{document}